\documentclass[sort,compress,11pt]{elsarticle}

\usepackage{amssymb}
\usepackage{amsthm}
\usepackage{amsmath, esint}
\usepackage{mathtools}
\usepackage{mathrsfs}
\usepackage[ruled, lined, linesnumbered, longend]{algorithm2e}
\usepackage{array}
\usepackage{booktabs}
\usepackage{multirow}
\usepackage{listings}
\usepackage{tabu}
\usepackage{enumerate}
\usepackage{lineno}
\usepackage{fullpage}
\usepackage{xcolor}
\usepackage[colorinlistoftodos]{todonotes}
\usepackage[colorlinks=true]{hyperref}
\usepackage{url}
\usepackage{textcomp}

\usepackage{graphicx}
\usepackage{subfig}

\usepackage{makecell}
\usepackage[export]{adjustbox}
\usepackage{caption}
\usetikzlibrary{shapes}
\biboptions{numbers,comma,round,square}
\usepackage[para,online,flushleft]{threeparttable}
\usepackage{orcidlink}

\journal{Elsevier}
\DontPrintSemicolon
\definecolor{orcidlogocol}{HTML}{A6CE39}
\begin{document}
\makeatletter
\def\ps@pprintTitle{%
  \let\@oddhead\@empty
  \let\@evenhead\@empty
  \let\@oddfoot\@empty
  \let\@evenfoot\@oddfoot
}
\makeatother
\begin{frontmatter}

\title{DiFVM: A Vectorized Graph-Based Finite Volume Solver for Differentiable CFD on Unstructured Meshes}

\author[ndAME]{Pan Du\fnref{equal}}
\author[ndAME,cornellMAE]{Yongqi Li\fnref{equal}}
\author[cornellMAE]{Mingqi Xu}
\author[ndAME,cornellMAE]{Jian-Xun Wang\orcidlink{0000-0002-9030-1733}\corref{corr}}

\fntext[equal]{These authors contributed equally to this work.}

\address[ndAME]{Department of Aerospace and Mechanical Engineering, University of Notre Dame, Notre Dame, IN}
\address[cornellMAE]{Sibley School of Mechanical and Aerospace Engineering, Cornell University, Ithaca, NY, USA}
%

\cortext[corr]{Corresponding author. Tel: +1 540 3156512}
\ead{jw2837@cornell.edu}

\begin{abstract}
Differentiable programming has emerged as a structural prerequisite for gradient-based inverse problems and end-to-end hybrid physics--machine learning in computational fluid dynamics. However, existing differentiable CFD platforms are confined to structured Cartesian grids, excluding the geometrically complex domains where body-conforming unstructured discretizations are indispensable.
We present \textbf{DiFVM}, the first GPU-accelerated, end-to-end differentiable finite-volume CFD solver operating natively on unstructured polyhedral meshes. The key enabling insight is a structural isomorphism between finite-volume discretization and graph neural network message-passing: by reformulating all FVM operators as static scatter/gather primitives on the mesh connectivity graph, DiFVM transforms irregular unstructured connectivity into a first-class GPU data structure. All operations are implemented in JAX/XLA, providing just-in-time compilation, operator fusion, and automatic differentiation through the complete simulation pipeline. Differentiable Windkessel outlet boundary conditions are provided for cardiovascular applications, and DiFVM accepts standard OpenFOAM case directories without modification for seamless adoption in existing workflows. Forward validation across benchmarks spanning canonical flows to patient-specific hemodynamics demonstrates close agreement with OpenFOAM, and end-to-end differentiability is demonstrated through inference of Windkessel parameters from sparse observations. DiFVM bridges the critical gap between differentiable programming and unstructured-mesh CFD, enabling gradient-based inverse problems and physics-integrated machine learning on complex engineering geometries.
\end{abstract}

\begin{keyword}
GPU-acceleration \sep Graph message passing \sep Inverse problems \sep Cardiovascular hemodynamics \sep Physics-integrated machine learning \sep JAX
\end{keyword}
\end{frontmatter}

\section{Introduction}

Computational fluid dynamics (CFD) has long served as an indispensable tool for modeling and understanding complex flow phenomena across science and engineering, ranging from aerodynamics and combustion to biological fluid systems and advanced manufacturing~\cite{fujii2005progress,jujjavarapu2024computational,zawawi2018review}. Decades of development have produced mature, high-fidelity solvers capable of simulating the Navier--Stokes equations on complex geometries at engineering-relevant scales~\cite{rojas2016review}. The canonical CFD workflow is straightforward in principle: given a computational geometry, prescribed boundary conditions, physical parameters, and model coefficients, one numerically solves the governing equations to obtain the flow field. While this \emph{forward-simulation} paradigm has been highly successful, its effectiveness rests on an assumption that is rarely satisfied in practice: that all model inputs are known, accurate, and sufficient to close the system. 
In most real engineering and biomedical problems, boundary conditions are often partially observed and must be inferred from indirect measurements; material or constitutive parameters may be spatially heterogeneous and poorly characterized; and turbulence closures introduce model-form uncertainty associated with unresolved physics that cannot be derived from first principles alone. These challenges are naturally formulated as \emph{PDE-constrained inverse problems}, in which unknown inputs, parameters, or closure terms are recovered by enforcing consistency between simulated outputs and available observations or design targets. Representative examples include recovering vascular resistance and compliance from sparse clinical pressure and flow measurements~\cite{pant2017inverse,wang2019data}, inferring aerodynamic surface geometry from pressure distributions in shape optimization~\cite{bui2004aerodynamic,li2019data}, and identifying turbulence closure contributions from experimental or high-fidelity reference data~\cite{lino2023current,wang2019prediction}. Critically, solving such problems at scale necessitates an outer optimization loop that repeatedly queries the forward solver and updates the unknowns based on gradient information. Scalable sensitivity computation via adjoint methods or automatic differentiation (AD) is essential, since finite-difference or derivative-free strategies become computationally prohibitive in the high-dimensional parameter spaces typical of CFD applications.

The need for gradient access has grown more acute as observational data has become abundant and the hybridization of ML with CFD has emerged as a dominant paradigm in scientific computing~\cite{brunton2020machine}. The emergence of large-scale experimental databases, high-resolution sensing arrays, and medical imaging has created new opportunities for \emph{data-augmented CFD}: workflows in which measurements actively shape or correct the simulation rather than merely validate it. Data assimilation methods fuse simulation outputs with observations to produce state estimates more accurate than either source alone~\cite{wang2016data,da2018ensemble}, but require gradients to propagate information from the measurement space back through the governing equations and into the model state~\cite{maulik2022efficient,wang2025variational}. 
The hybridization of ML with CFD represents a further and more ambitious step in this direction: rather than merely assimilating data into an existing simulation state, ML components are trained 
to learn the unknown or unresolved physics embedded within the governing equations themselves. Early hybrid approaches adopted \emph{loosely coupled} architectures: ML models were trained offline using high-fidelity simulation data and subsequently embedded into conventional CFD solvers as closure models~\cite{pan2018data,zhuang2021learned}, particularly for turbulence modeling, where data-driven subgrid-scale closures and Reynolds stress corrections have been extensively explored~\cite{duraisamy2019turbulence,wang2017physics,yang2019predictive,akolekar2019development,prakash2024invariant}. While effective in certain regimes, these offline-trained models are fundamentally not CFD-consistent. This is because training is decoupled from the solver, and the learned closures have no guarantee of stability or accuracy when deployed in the forward simulation loop~\cite{wu2019reynolds}. This manifests concretely as \emph{a posteriori} degradation, i.e., models that perform well in \emph{a priori} evaluation against reference data but exhibit significant errors or numerical instability when integrated into the live solver~\cite{beck2019deep,sirignano2020dpm,macart2021embedded,fang2023toward}. The root cause is the severed feedback loop - without gradients flowing through the solver, the ML model cannot learn to correct for the accumulated errors encountered during deployment~\cite{fan2024differentiable}.

This limitation has motivated a shift toward \emph{fully coupled} hybrid physics--ML frameworks via differentiable programming~\cite{sapienza2024differentiable}, in which ML components and CFD  operators are jointly optimized end-to-end within a single differentiable computational graph. When gradients propagate \emph{through} the solver through flux computations, pressure solves, and time integration steps, the neural network learns representations that are consistent with the numerical solver's dynamics, not merely with offline training data. This paradigm has produced substantial advances in solver-in-the-loop learned turbulence closures~\cite{list2022learned,shankar2025differentiable}, data-driven discretizations and end-to-end solver corrections~\cite{bar2019learning,um2020solver,kochkov2021machine}, and hybrid differentiable solvers with learnable numerical components (e.g., fluxes or coarse-model augmentation)~\cite{bezgin2021data,belbute2020combining}. Crucially, differentiable hybrid modeling goes beyond ML embedded \emph{within} a CFD solver: it enables PDE-integrated neural architectures where the CFD solver or its discretized operators are embedded \emph{as structured layers within} an advanced neural architecture~\cite{liu2024multi}. Under this \emph{neural differentiable modeling} paradigm~\cite{akhare2023diffhybrid,akhare2025hybridndiff,akhare2025implicit}, discretized PDE operators serve as fixed, physics-encoding layers interlaced with trainable components such as ConvLSTM blocks, graph kernels, or residual networks, forming physics-integrated surrogate models of greater physical consistency and data efficiency than purely black-box alternatives. Wang and co-workers have demonstrated this direction across multiple domains, including turbulence modeling~\cite{fan2025neural}, fluid--structure interaction (FSI)~\cite{fan2024differentiable}, and multiscale thermal transport~\cite{shang2025jax,akhare2024probabilistic}. For example, in hybrid neural differentiable solvers for turbulence and FSI, discretized Navier--Stokes operators are encoded as fixed convolutional layers within a recurrent architecture and interleaved with trainable ConvLSTM blocks and a diffusion generative module that learn subgrid-scale corrections and generate small-scale turbulent structures, forming a physics-integrated neural network rather than a neural-network-augmented solver~\cite{fan2025neural,fan2024differentiable}. Across all of these settings, the gradient path through the physics operators is what enables physically consistent learning from sparse and indirect data. \emph{Differentiability through the solver is therefore not a convenience but a structural prerequisite for the hybrid physics--ML architectures that the field is converging toward.}

Despite the clarity of this requirement, conventional CFD solvers cannot meet it. Most production-grade codes are implemented in C++ or Fortran, structured around explicit loops over irregular mesh connectivity, and optimized for CPU parallelism via MPI~\cite{rojas2016review}. Gradient access in these systems has historically been achieved through adjoint methods (e.g., SU2~\cite{economon2016su2}, DOLFIN-adjoint~\cite{mitusch2019dolfin}, ADjoint~\cite{mader2008adjoint}, STAR-CCM+~\cite{ccmSu}), but adjoint implementations are code-intrusive, solver-specific, and require substantial re-derivation for each new problem configuration. More critically, they are architecturally decoupled from the AD ecosystems such as JAX~\cite{jax2018github}, PyTorch~\cite{paszke2019pytorch}, TensorFlow~\cite{abadi2016tensorflow}, Julia~\cite{Julia-2017}, where modern neural network components are built and trained. This decoupling prevents the seamless composition of physics operators with learnable modules within a single differentiable computational graph, which is precisely what hybrid physics--ML frameworks require. Beyond differentiability, conventional solvers carry a second structural incompatibility: they are CPU-native, relying on loop-based execution over irregular connectivity that maps poorly onto the single instruction multiple 
data (SIMD) and single instruction multiple threads (SIMT) execution models of modern GPUs and TPUs~\cite{posey2022gpu}. Even where adjoint gradients are available, the underlying solver 
remains orders of magnitude slower than GPU-native ML infrastructure, making iterative gradient-driven workflows computationally impractical at modern application scales.

These limitations have motivated a wave of GPU-native differentiable CFD platforms implemented within AD-friendly programming frameworks~\cite{kochkov2021machine,holl2024bf,weymouth2024waterlily,bezgin2025jax,franz2025pict, fan2026diff, shang2025jax}. Representative frameworks include JAX-CFD~\cite{kochkov2021machine}, a JAX-native incompressible flow solver on structured Cartesian meshes widely adopted in hybrid ML--CFD studies; PhiFlow~\cite{holl2024bf}, a backend-agnostic differentiable PDE framework for rapid prototyping of 
gradient-based optimization with simplified fluid models; JAX-Fluids~\cite{bezgin2025jax}, a high-order finite-volume solver for compressible and multiphase flows with multi-GPU scalability, but whose compressible-flow focus and structured Cartesian topology render it 
unsuited for the low-Mach incompressible regimes central to cardiovascular, FSI, and most industrial applications; WaterLily~\cite{weymouth2024waterlily}, which handles flows around moving bodies via a Cartesian immersed boundary method but lacks turbulence closure models and two-way structural coupling; and PICT~\cite{franz2025pict}, which demonstrates gradient-based inverse modeling in transient dynamics but targets simplified physics on structured discretizations. A significant step forward is represented by Diff-FlowFSI~\cite{fan2026diff}, the first differentiable CFD 
platform designed for engineering-scale turbulence and strongly coupled multiphysics FSI. Unlike prior frameworks, Diff-FlowFSI supports subgrid-scale and wall models for wall-bounded turbulence and strong two-way coupling between the incompressible Navier--Stokes equations and elastic structural dynamics, all within a fully JIT-compiled, vectorized JAX implementation that achieves up to two orders of magnitude speedup over CPU-parallelized CFD solvers while maintaining 
end-to-end differentiability through the complete simulation pipeline. Collectively, these platforms have established GPU-native differentiable simulation as a practical substrate for gradient-driven workflows across an expanding range of physical systems.

Despite this progress, all existing differentiable CFD frameworks share a fundamental and unresolved limitation: they are confined to structured Cartesian grids. This restriction reflects a genuine 
algorithmic challenge: unstructured meshes exhibit irregular cell connectivity, variable face-neighbor counts, geometric non-orthogonality, and non-coalesced memory access patterns that are 
fundamentally hostile to the vectorized SIMD/SIMT execution model of modern GPU hardware. Yet unstructured meshes are indispensable for the geometrically complex domains that define the most consequential CFD applications (e.g., patient-specific vascular anatomies, turbomachinery passages, aircraft surfaces, and general engineering geometries with irregular boundaries) where body-conforming discretizations are required for accurate boundary layer resolution and physically consistent flux computation. The FVM is the natural discretization for such geometries, offering a geometry-agnostic formulation, strict local conservation, and robust handling of arbitrary cell topologies and boundary conditions. The absence of a GPU-native, end-to-end differentiable FVM solver on unstructured meshes therefore represents not a peripheral gap but a fundamental bottleneck that limits the geometric reach of the entire differentiable CFD ecosystem and, by extension, the hybrid physics--ML frameworks built upon it. \emph{To the best of our knowledge, no existing framework provides a GPU-efficient, fully vectorized, end-to-end differentiable finite-volume CFD solver that operates natively on 3D unstructured meshes.}

In this work, we bridge this gap by presenting \textbf{DiFVM}, a first-of-its-kind GPU-accelerated, vectorized, end-to-end differentiable finite-volume CFD solver that operates natively on unstructured polyhedral meshes. The key enabling insight is a structural isomorphism between FVM discretization and the message-passing paradigm of graph neural networks (GNNs): computing inter-cell fluxes from cell-center quantities and aggregating them over each control volume is precisely the local-computation-and-neighborhood-aggregation pattern that GNN frameworks already execute in a fully vectorized, GPU-native manner via scatter/gather primitives~\cite{pfaff2020learning}. By 
reformulating all FVM operators as static graph-based message-passing primitives, the irregular connectivity of unstructured meshes is transformed from a computational liability into a first-class data structure amenable to GPU vectorization, with JAX's XLA compiler providing just-in-time 
compilation, operator fusion, and automatic differentiation through the complete pipeline. DiFVM solves the incompressible Navier--Stokes equations with all operators formulated in this graph-based framework and temporal integration unrolled via \texttt{jax.lax.scan}. It accepts any unstructured polyhedral mesh, including pure tetrahedral, hexahedral, prismatic, and arbitrary hybrid topologies, 
and is directly compatible with OpenFOAM case directories: an existing OpenFOAM case can be passed to DiFVM without modification, enabling seamless adoption within established simulation workflows. For cardiovascular applications, three-element Windkessel (RCR) outlet boundary conditions are additionally implemented, supporting patient-specific hemodynamic simulation and end-to-end parameter inference from sparse clinical measurements.
The main contributions of this work are:
\begin{itemize}
    \item \textbf{First GPU-native vectorized differentiable FVM solver on unstructured meshes.} DiFVM supports fully unstructured polyhedral meshes with arbitrary element types, including tetrahedra, hexahedra, prisms, pyramids, and hybrid combinations, together with non-orthogonality correction and complex boundary conditions. These capabilities are absent from all existing differentiable CFD frameworks.

    \item \textbf{Graph-based vectorization of finite-volume operators.} Convective fluxes, diffusive fluxes, and the PISO pressure--velocity coupling are reformulated as static message-passing graph primitives and compiled to fused XLA GPU kernels, achieving substantial speedups over CPU-parallelized OpenFOAM on large-scale transient benchmarks.

    \item \textbf{Drop-in OpenFOAM compatibility.} DiFVM directly accepts OpenFOAM case directories without modification, enabling immediate adoption in existing workflows and facilitating 
    quantitative comparison under identical numerical settings.

    \item \textbf{Differentiable Windkessel boundary conditions for cardiovascular applications.} DiFVM implements three-element Windkessel (RCR) outlet boundary conditions within the differentiable pipeline, enabling patient-specific hemodynamic simulation on anatomically realistic vascular geometries. 

    \item \textbf{Comprehensive forward validation across multiple benchmarks.} Accuracy is demonstrated on a 3D Poisson problem, passive scalar advection, lid-driven cavity flow, elbow flow, cylinder vortex shedding, and patient-specific multi-branch aortic flow, with quantitative comparison against OpenFOAM~\cite{weller1998tensorial} and SimVascular~\cite{updegrove2017simvascular}.

    \item \textbf{End-to-end automatic differentiation validated on inverse problems.} The AD pipeline is verified through gradient-based recovery of lid boundary conditions in cavity flow 
    and simultaneous inference of six Windkessel parameters from sparse pressure and flow-rate observations in a pulsatile vascular flow.
\end{itemize}

The remainder of this paper is organized as follows. Section~\ref{sec:methods} describes the DiFVM methodology, including finite-volume discretization, solver schemes, and GPU-oriented vectorization strategies. Section~\ref{sec:results} presents benchmark validation, efficiency comparisons, and inverse problem demonstrations. Section~\ref{sec:discussion} discusses current limitations and future directions. Section~\ref{sec:conclusion} concludes the paper.

\section{Methodology}
\label{sec:methods}

\subsection{Overview}

\begin{figure}[htb!]
\centering
\includegraphics[width=\linewidth]{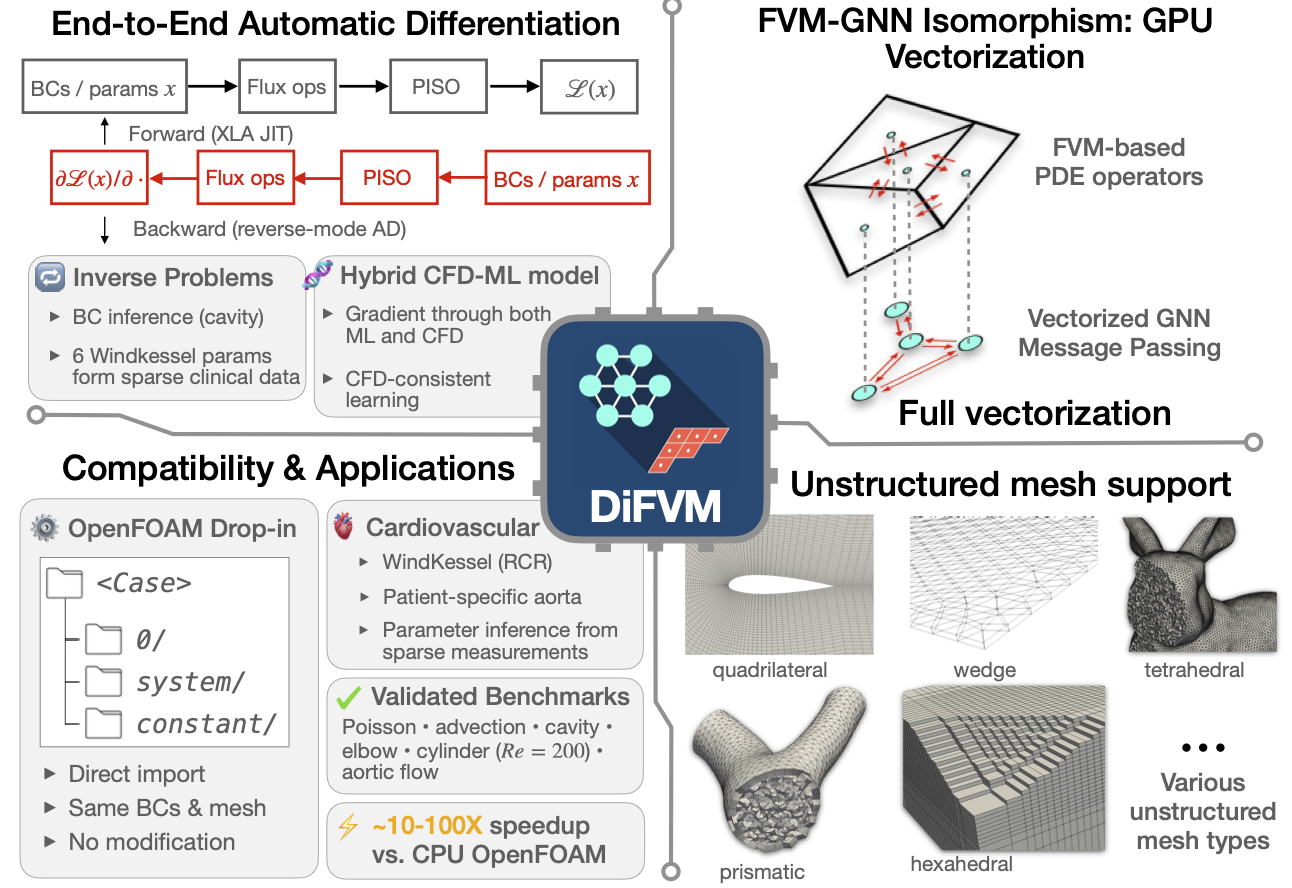} 
\caption{Overview of DiFVM: end-to-end differentiable graph-based FVM solver with GPU vectorization, native unstructured mesh support, and OpenFOAM drop-in compatibility, supporting inverse and hybrid ML-CFD modeling.}
\label{fig:DiFVM}
\end{figure}

DiFVM is designed as a fully differentiable finite-volume CFD solver that addresses the fundamental absence of GPU-native, end-to-end differentiable simulation on unstructured meshes.  As illustrated in
Figure~\ref{fig:DiFVM}, DiFVM is built around four tightly integrated capabilities: \emph{(i)} native support for arbitrary unstructured polyhedral meshes; \emph{(ii)} full GPU vectorization via a reformulation of all FVM operators as graph-based message-passing primitives; \emph{(iii)} end-to-end automatic differentiation through the complete simulation pipeline, from boundary conditions and physical parameters through every flux computation, linear solve, and time integration step; and \emph{(iv)} drop-in OpenFOAM compatibility with differentiable Windkessel boundary conditions for cardiovascular applications.  All numerical operations are implemented natively in the JAX/XLA ecosystem, enabling just-in-time (JIT) compilation, operator fusion, and GPU-accelerated execution that achieves one to two orders of magnitude speedup over CPU-parallelized CFD solvers.

The central design insight is a structural isomorphism between FVM discretization and the message-passing (MP) paradigm of GNNs~\cite{pfaff2020learning}: computing inter-cell fluxes from cell-center quantities and aggregating them over each control volume is precisely the MP (message construction and neighborhood aggregation) pattern that GNN frameworks already execute via scatter/gather primitives in a fully vectorized, GPU-native manner. Critically, this mapping is exact and lossless: it is a reformulation of the classical FVM operators, not an approximation.  By expressing all FVM operators as static graph-based MP primitives over the mesh connectivity graph, we represent unstructured-mesh neighbor relations explicitly and compute fluxes via batched gather-compute-scatter kernels, enabling efficient GPU vectorization despite irregular connectivity. JAX's Accelerated Linear Algebra (XLA) compiler provides JIT compilation and operator fusion across the complete solver pipeline, and temporal integration is unrolled via \texttt{jax.lax.scan} to eliminate Python-side overhead and enable memory-efficient backpropagation through time.

\subsection{Governing Equations}
\label{sec:governing}

DiFVM solves the unsteady incompressible Navier--Stokes equations for Newtonian fluids on a bounded domain $\Omega \subset \mathbb{R}^3$ with boundary $\partial\Omega = \Gamma_D \cup \Gamma_N \cup \Gamma_W$:
\begin{align}
  \nabla \cdot \mathbf{u} &= 0,
  \label{eq:continuity} \\
  \frac{\partial \mathbf{u}}{\partial t}
  + \nabla \cdot (\mathbf{u} \otimes \mathbf{u})
  &= -\frac{1}{\rho}\nabla p
  + \nabla \cdot (\nu \nabla \mathbf{u})
  + \mathbf{f},
  \label{eq:momentum}
\end{align}
where $\mathbf{u}(\mathbf{x},t) \in \mathbb{R}^{3}$ is the velocity field, $p(\mathbf{x},t)$ is the pressure field, $\nu$ is the kinematic viscosity, $\mathbf{f}$ denotes body forces, and $\otimes$ denotes the outer product. The continuity equation~\eqref{eq:continuity} enforces the divergence-free constraint, and the momentum equation~\eqref{eq:momentum} balances inertia, pressure gradient, viscous diffusion, and body forces. The system is closed by initial and boundary conditions:
\begin{equation}
  \mathbf{u}(\mathbf{x}, 0) = \mathbf{u}_0(\mathbf{x}),
  \quad \mathbf{x} \in \Omega,
  \label{eq:ic}
\end{equation}
\begin{alignat}{3}
  \mathbf{u} &= \mathbf{u}_{\Gamma}(\mathbf{x},t),
  &\quad &\mathbf{x} \in \Gamma_D
  \quad &(\text{Dirichlet: inlet, wall, or moving surface}),
  \label{eq:bc_dirichlet} \\
  -p \mathbf{n} + \nu(\nabla\mathbf{u})\mathbf{n}
  &= \mathbf{h}(\mathbf{x},t),
  &\quad &\mathbf{x} \in \Gamma_N
  \quad &(\text{Neumann: outlet or traction}),
  \label{eq:bc_neumann} \\
  \mathbf{u} &= \mathbf{0},
  &\quad &\mathbf{x} \in \Gamma_W
  \quad &(\text{no-slip wall}),
  \label{eq:bc_wall}
\end{alignat}
where $\mathbf{n}$ denotes the outward unit normal on $\partial\Omega$. For cardiovascular applications, the Neumann outlet condition $\mathbf{h}$ is prescribed by the Windkessel model detailed in Section~\ref{sec:windkessel}, which dynamically couples the outlet pressure to the instantaneous flow rate through a lumped-parameter vascular resistance--compliance circuit.

\subsection{Finite-Volume Discretization on Unstructured Meshes}
\label{sec:fvm}
\subsubsection{Mesh data structure and collocated cell-centered layout}

The computational domain $\Omega$ is discretized into a set of non-overlapping polyhedral control volumes (cells) $\{\Omega_i\}$, $i = 1, \ldots, N_c$, where $N_c$ is the total number of cells.  Each cell $\Omega_i$ is bounded by a set of faces $\mathcal{F}(i) = \{f_1, f_2, \ldots\}$, where each internal face $f$
is shared between exactly two cells (owner $i$ and neighbor $j$), and each boundary face is associated with a single cell and a prescribed boundary condition. For every face $f$, the signed area vector $\mathbf{S}_f$ points from the owner cell outward, with magnitude equal to the face area $A_f = |\mathbf{S}_f|$ and direction equal to the outward unit normal $\hat{\mathbf{n}}_f$.  The centroid of cell $i$ is denoted $\mathbf{x}_i$, the cell volume is $V_i$, and the face centroid is $\mathbf{x}_f$.

DiFVM employs a \emph{collocated cell-centered} arrangement: both the velocity $\mathbf{u}_i$ and the pressure $p_i$ are stored at cell centroids. This arrangement naturally accommodates arbitrary polyhedral
elements and simplifies the graph-based vectorization strategy described in Section~\ref{sec:mp}. To prevent the spurious pressure--velocity decoupling that arises on collocated grids, Rhie--Chow momentum interpolation~\cite{rhie1983numerical} is applied when constructing the face velocity for the continuity equation, as detailed in Section~\ref{sec:piso_graph}.
The complete mesh topology (i.e., cell centroids, face area vectors, owner/neighbor cell indices, and boundary patch assignments) is extracted at initialization and stored as static, immutable arrays. Because the mesh geometry is fixed throughout the simulation, the resulting connectivity graph constitutes a static data structure that can be precomputed once, loaded into GPU memory, and reused across all time steps without modification. 

\subsubsection{Spatial discretization of convective and diffusive fluxes}
\label{sec:fvm-flux}
Integrating the momentum equation~\eqref{eq:momentum} over cell $\Omega_i$ and applying the divergence theorem yields the semi-discrete FVM balance:
\begin{equation}
  V_i \frac{d\mathbf{u}_i}{dt}
  = -\sum_{f \in \mathcal{F}(i)}
      \underbrace{(\mathbf{u} \otimes \mathbf{u})_f}_{
        \mathbf{J}^C_f} \cdot \mathbf{S}_f
  + \sum_{f \in \mathcal{F}(i)}
      \underbrace{\nu\,(\nabla \mathbf{u})_f}_{
        \mathbf{J}^D_f} \cdot \mathbf{S}_f
  - \frac{1}{\rho}\sum_{f \in \mathcal{F}(i)} p_f \,\mathbf{S}_f
  + V_i \mathbf{f},
  \label{eq:fvm_momentum}
\end{equation}
where the convective flux $\mathbf{J}^C = \mathbf{u}\otimes\mathbf{u}$ and diffusive flux $\mathbf{J}^D = \nu\nabla\mathbf{u}$ are second-order tensors evaluated at each face centroid, and $\mathbf{J}^C_f\cdot\mathbf{S}_f$, $\mathbf{J}^D_f\cdot\mathbf{S}_f$ denote the net convective and diffusive momentum fluxes through face $f$, respectively. The convective face flux simplifies to $\mathbf{J}^C_f\cdot\mathbf{S}_f = (\mathbf{u}_f\cdot\mathbf{S}_f)\,\mathbf{u}_f$, which is the form used in the discrete implementation. The face area vector $\mathbf{S}_f$ points outward from $\Omega_i$, and $p_f$, $\mathbf{u}_f$ denote the face-interpolated pressure and velocity. The discrete treatment of convective flux and diffusive flux differs fundamentally and is detailed separately as follows.

\paragraph{Convective flux discretization}

The convective flux tensor $\mathbf{J}^C = \mathbf{u}\otimes\mathbf{u}$ is a rank-2 tensor, but its dot product with the face area vector reduces to a scalar-transport problem for each velocity component. Specifically, applying the identity $(\mathbf{a}\otimes\mathbf{b})\cdot\mathbf{n} = (\mathbf{b}\cdot\mathbf{n})\,\mathbf{a}$:
\begin{equation}
  \mathbf{J}^C_f \cdot \mathbf{S}_f
  = (\mathbf{u}\otimes\mathbf{u})_f\cdot\mathbf{S}_f
  = \dot{m}_f\,\mathbf{u}_f,
  \qquad
  \dot{m}_f = \mathbf{u}_f \cdot \mathbf{S}_f,
  \label{eq:conv_face_flux}
\end{equation}
where $\dot{m}_f$ is the face-normal volumetric flux through face $f$. For a generic transported scalar $\phi$ (representing any single velocity component), the convective contribution at face $f$ is therefore $\dot{m}_f\,\phi_f$, and the problem of evaluating $\mathbf{J}^C_f\cdot\mathbf{S}_f$ reduces entirely to reconstructing the face value $\phi_f$ from the two adjacent cell-center values $\phi_i$ (owner) and $\phi_j$ (neighbor).

DiFVM supports multiple face reconstruction schemes. The \emph{first-order upwind} (U) scheme selects the upwind cell value based on the sign of $\dot{m}_f$:
\begin{equation}
  \phi_f^U
  = \begin{cases}
      \phi_i & \dot{m}_f \geq 0 \quad (\text{flow from } i \to j), \\
      \phi_j & \dot{m}_f < 0  \quad (\text{flow from } j \to i),
    \end{cases}
  \label{eq:upwind}
\end{equation}
which is bounded and unconditionally stable but only first-order accurate. To recover higher-order accuracy while retaining numerical stability, DiFVM employs a \emph{deferred correction} strategy:
\begin{equation}
  \dot{m}_f\,\phi_f^{HO}
  = \underbrace{\dot{m}_f\,\phi_f^{U}}_{\text{implicit}}
  + \underbrace{\dot{m}_f\!\left(\phi_f^{HO} - \phi_f^{U}\right)}_{\text{explicit correction}},
  \label{eq:deferred_correction}
\end{equation}
where superscripts $U$ and $HO$ denote the first-order upwind and higher-order reconstructions, respectively. This preserves the diagonal dominance and sparsity structure of the momentum matrix while achieving higher-order accuracy at convergence. The higher-order face value $\phi_f^{HO}$ is obtained via the \emph{second-order upwind} (SOU) scheme, which reconstructs $\phi_f$ using the cell-center value and gradient at the upwind cell:
\begin{equation}
  \phi_f^{HO} = \phi_C + (\nabla\phi)_C \cdot \mathbf{d}_{Cf},
  \qquad
  \mathbf{d}_{Cf} = \mathbf{x}_f - \mathbf{x}_C,
  \label{eq:sou}
\end{equation}
where $C$ denotes the upwind cell (owner $i$ or neighbor $j$ depending on the sign of $\dot{m}_f$) and $(\nabla\phi)_C$ is the cell-centered gradient evaluated via the Gauss--Green theorem:
\begin{equation}
	(\nabla\phi)_i
	= \frac{1}{V_i}
	\sum_{f \in \mathcal{F}(i)} \phi_f\,\mathbf{S}_f,
	\label{eq:gauss_green}
\end{equation}
with $\phi_f$ obtained by linear interpolation.

\paragraph{Diffusive flux discretization and non-orthogonality correction}
The diffusive flux tensor $\mathbf{J}^D = \nu\nabla\mathbf{u}$ is a rank-2 tensor.  Its face flux $\mathbf{J}^D_f\cdot\mathbf{S}_f$, however, decouples into independent scalar problems for each velocity
component.  Writing $(\nabla\mathbf{u})_{kl} = \partial u^k/\partial x^l$ and contracting with $\mathbf{S}_f$:
\begin{equation}
	\bigl[\mathbf{J}^D_f \cdot \mathbf{S}_f\bigr]^k
	= \nu\,(\nabla u^k)_f \cdot \mathbf{S}_f,
	\label{eq:diff_component}
\end{equation}
so the problem reduces, for each velocity component $\phi \equiv u^k$, to evaluating the scalar face flux $\nu\,(\nabla\phi)_f\cdot\mathbf{S}_f$. This is structurally analogous to the convective case, but the challenge here is not face-value reconstruction but accurate evaluation of the \emph{face-normal gradient}
$(\nabla\phi)_f\cdot\mathbf{S}_f$.

On orthogonal meshes ($\mathbf{S}_f \parallel \mathbf{d}_{ij}$), the face-normal gradient reduces to a compact two-point stencil:
\begin{equation}
	\nu\,(\nabla\phi)_f\cdot\mathbf{S}_f
	\approx \nu\,|\mathbf{S}_f|\,
	\frac{\phi_j - \phi_i}{|\mathbf{d}_{ij}|},
	\label{eq:diff_ortho}
\end{equation}
which is second-order accurate. 
On general unstructured meshes, $\mathbf{S}_f$ and $\mathbf{d}_{ij}$ are not parallel, and equation~\eqref{eq:diff_ortho} incurs a first-order truncation error.  To restore second-order accuracy,
$\mathbf{S}_f$ is decomposed into an orthogonal component $\mathbf{\Delta}_f$ and a non-orthogonal correction
$\mathbf{k}_f = \mathbf{S}_f - \mathbf{\Delta}_f$:
\begin{equation}
  \nu\,(\nabla\phi)_f\cdot\mathbf{S}_f
  = \underbrace{
      \nu\,|\mathbf{\Delta}_f|\,
      \frac{\phi_j - \phi_i}{|\mathbf{d}_{ij}|}
    }_{\text{orthogonal}}
  + \underbrace{
      \nu\,(\nabla\phi)_f \cdot \mathbf{k}_f
    }_{\text{non-orthogonal correction}},
  \label{eq:nonortho_flux}
\end{equation}
where the choice of $\mathbf{\Delta}_f$ defines the correction scheme~\cite{jasak1996error}. With $\hat{\mathbf{d}}_{ij} = \mathbf{d}_{ij}/|\mathbf{d}_{ij}|$ and $\hat{\mathbf{S}}_f = \mathbf{S}_f/|\mathbf{S}_f|$ being the unit vectors along the cell-to-cell direction and face normal,
respectively, the \emph{minimum correction} approach sets $\mathbf{\Delta}_f = (\mathbf{S}_f\cdot\hat{\mathbf{d}}_{ij})
\hat{\mathbf{d}}_{ij}$, minimizing $|\mathbf{k}_f|$ but underperforming on highly non-orthogonal meshes; the
\emph{orthogonal correction} approach sets $\mathbf{\Delta}_f = |\mathbf{S}_f|\hat{\mathbf{d}}_{ij}$, which can overestimate the correction on skewed meshes; and the \emph{over-relaxed} approach, adopted in DiFVM, sets $\mathbf{\Delta}_f = \bigl(|\mathbf{S}_f|/|\hat{\mathbf{S}}_f\cdot \hat{\mathbf{d}}_{ij}|\bigr)\hat{\mathbf{d}}_{ij}$, which provides the best balance between accuracy and stability for the moderately non-orthogonal meshes typical of complex engineering geometries~\cite{jasak1996error}. The cell-centered gradient $(\nabla\phi)_i$ required in the correction term is evaluated via the Gauss--Green theorem.

\subsection{Graph-Based Reformulation of the FVM Solver for GPU Vectorization}
\label{sec:mp}

\subsubsection{Structural isomorphism between FVM and GNN and static graph construction}
\label{sec:isomorphism}
\begin{figure}[htb!]
\centering
\includegraphics[width=\linewidth]{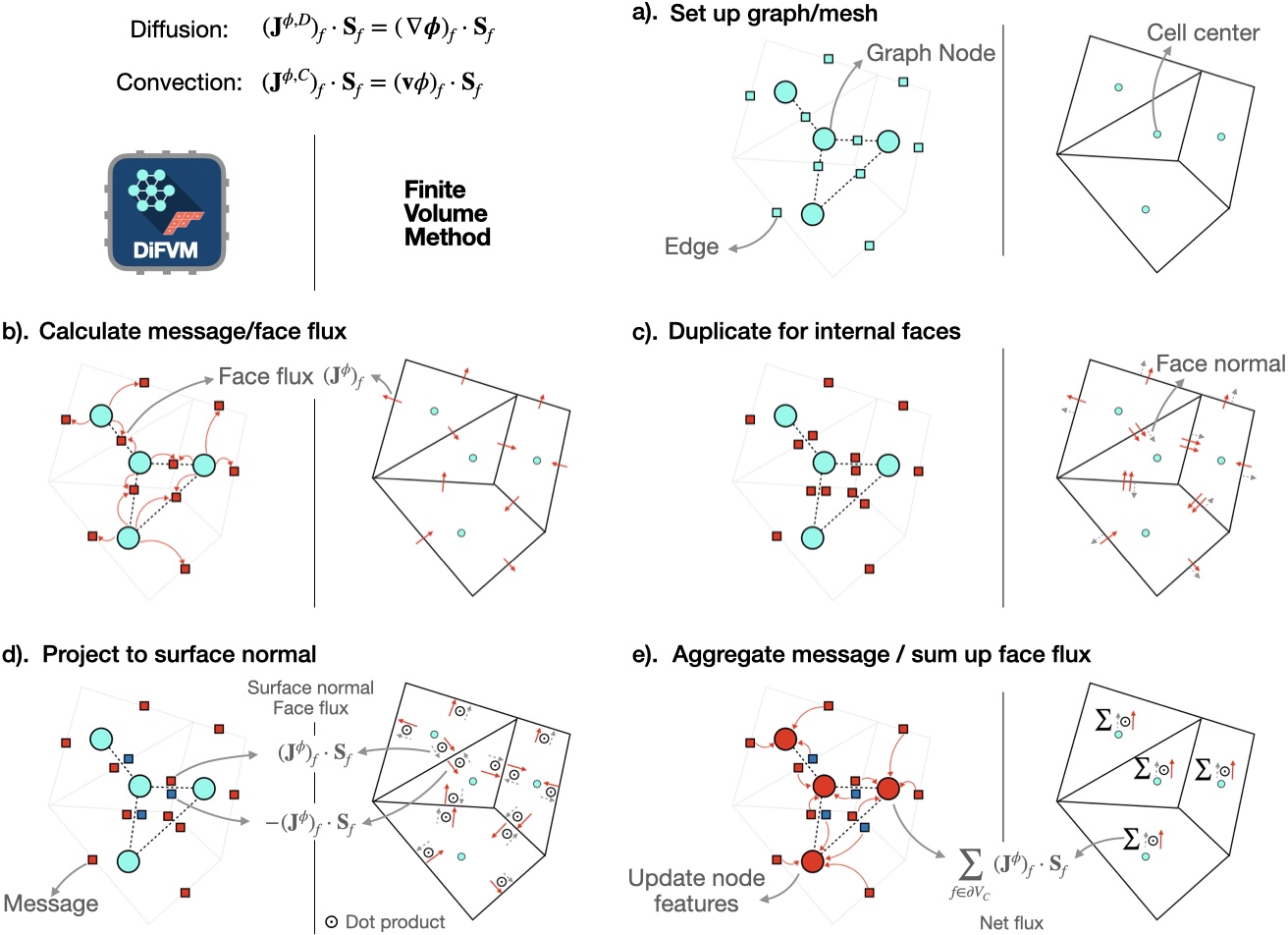}
\caption{Analogy between graph message passing and finite-volume discretization on unstructured meshes.  Cell centers and shared faces map to graph nodes and directed edges, respectively. Face flux
computation and aggregation map exactly onto the message-construction and node-update steps of a GNN, enabling full GPU vectorization via scatter/gather primitives.}
\label{fig:vectorization}
\end{figure}

The key insight enabling GPU vectorization on unstructured meshes is a structural isomorphism between FVM discretization and the message-passing (MP) paradigm of GNNs, illustrated in Figure~\ref{fig:vectorization}. As shown in the figure, the mesh is cast as a directed graph $\mathcal{G} = (\mathcal{V}, \mathcal{E})$, 
where each node $v_i \in \mathcal{V}$ corresponds to a cell centroid storing the state $(\mathbf{u}_i, p_i)$, and each directed edge $e_{ji} \in \mathcal{E}$ corresponds to an internal face, carrying information from neighbor cell $j$ to owner cell $i$. Boundary faces are treated as directed edges from ghost nodes to interior cells. Each edge carries static geometric attributes $\mathbf{g}_{ji} = (\mathbf{S}_f,\mathbf{x}_f)$ precomputed at initialization and fixed throughout the simulation.

At initialization, DiFVM constructs the following static arrays from the mesh, uploaded to GPU memory once and reused throughout the simulation: \texttt{owner}, \texttt{neighbor} $\in \mathbb{Z}^{|\mathcal{E}|}$ (owner/neighbor cell indices); \texttt{Sf}, \texttt{xf} $\in \mathbb{R}^{|\mathcal{E}|\times 3}$ (face area vectors, face centroids); \texttt{cf}, \texttt{vol} $\in \mathbb{R}^{N_c}$ (cell center, cell volumes); and boundary patch index arrays mapping each boundary face to its condition type and parameters. The dynamic state consists of only two node-level arrays: \texttt{U} $\in \mathbb{R}^{N_c\times 3}$ and \texttt{p} $\in \mathbb{R}^{N_c}$. Because all geometric arrays are static, the XLA compiler treats them as compile-time constants and fuses their access into optimized GPU kernels.

Under this graph representation, the FVM residual~\eqref{eq:fvm_momentum} maps exactly onto a three-stage MP operation ($\mathcal{M}$, $\mathcal{A}$, and $\mathcal{U}$ are message construction, aggregation, and update operators):
\begin{enumerate}
  \item \textbf{Message construction.} For each directed edge $e_{ji}$, compute the face flux message $\mathbf{m}_{ji}$ from the states of the two endpoint cells and the edge geometric attributes:
    \begin{equation}
      \mathbf{m}_{ji} 
      = \mathcal{M}(\mathbf{u}_i, \mathbf{u}_j, \mathbf{g}_{ji})
      = \underbrace{
          -\mathbf{J}^C_f\cdot\mathbf{S}_f
        }_{\text{convective}}
      + \underbrace{
          \mathbf{J}^D_f\cdot\mathbf{S}_f
        }_{\text{diffusive}}
      - \underbrace{
          \frac{1}{\rho}\,p_f\,\mathbf{S}_f
        }_{\text{pressure}},
      \label{eq:message}
    \end{equation}
    where $\mathbf{J}^C_f$ and $\mathbf{J}^D_f$ are the convective and diffusive flux tensors evaluated at face $f$.  Each message is a pointwise function of node and edge attributes with no dependence on the global mesh topology, and is therefore embarrassingly parallel across all edges.

  \item \textbf{Flux aggregation.} For each node $v_i$, aggregate all incoming edge messages:
    \begin{equation}
      \mathbf{R}_i
      = \mathcal{A}(\mathbf{m}_{ji}, \mathcal{E})
      = \sum_{j \in \mathcal{N}(i)} \sigma_{ij}\,\mathbf{m}_{ji},
      \label{eq:aggregate}
    \end{equation}
    where $\mathcal{N}(i)$ denotes the set of cells sharing a face with cell $i$, and $\sigma_{ij} = \pm 1$ encodes the outward orientation convention. This aggregation is implemented via \texttt{jax.ops.segment\_sum} over the edge-wise flux array indexed by the precomputed \texttt{owner} array, which is a single parallel scatter-add over all $N_c$ nodes.

  \item \textbf{Node update.} Update the cell state using the
    aggregated residual:
    \begin{equation}
      \mathbf{u}_i^{n+1} = \mathcal{U}(\mathbf{u}_i^n, \frac{\mathbf{R}_i}{V_i})
      \label{eq:node_update}
    \end{equation}
    where $V_i$ is the cell volume and $\mathcal{U}$ denotes the time integration operator (e.g.,
    forward Euler, $\mathbf{u}_i^{n+1} = \mathbf{u}_i^n + \Delta t\,\mathbf{R}_i/V_i$). For incompressible flows, $\mathcal{U}$ must enforce the divergence-free constraint $\nabla\cdot\mathbf{u}=0$, which cannot be achieved by a simple explicit step. DiFVM realizes $\mathcal{U}$ as a \emph{divergence-free constrained message-passing} procedure, detailed in Section~\ref{sec:piso_graph}.
\end{enumerate}
The irregular mesh connectivity, which in a conventional CPU solver requires sequential cell-by-cell traversal, is entirely encoded in the precomputed edge index arrays, which serve as static gather/scatter indices for batched tensor operations. The result is a fully vectorized residual computation reducible to two GPU kernels: one parallel gather--compute over all $|\mathcal{E}|$ edges (MP), and one parallel scatter-add over all $N_c$ nodes (flux aggregation), with no Python-level loops over mesh elements.

\subsubsection{PISO as divergence-free constrained message passing}
\label{sec:piso_graph}

A naive explicit node update~\eqref{eq:node_update} does not enforce $\nabla\cdot\mathbf{u}=0$ on a collocated grid, because velocity and pressure are not independently determined. The Pressure Implicit with Splitting of Operators (PISO) algorithm~\cite{issa1986solution} resolves this by decomposing $\mathcal{U}$ into three graph operations that together project the velocity onto the divergence-free subspace. Each operation is expressed exclusively in terms of scatter/gather primitives on $\mathcal{G}$, with no operations outside the graph abstraction.

\paragraph{Layer 1: Momentum MP layer}
Compute a predicted velocity $\mathbf{u}^*$ by advancing the momentum residual~\eqref{eq:fvm_momentum} from time $t^n$, treating the pressure gradient explicitly from $p^n$:
\begin{equation}
\frac{\mathbf{u}_i^* - \mathbf{u}_i^n}{\Delta t} = \frac{1}{V_i}\bigl[\mathbf{R}_i^{\mathrm{conv}}(\mathbf{u}^n) + \mathbf{R}_i^{\mathrm{diff}}(\mathbf{u}^n) + \mathbf{R}_i^{\mathrm{pres}}(p^n)\bigr], \label{eq:momentum_predictor}
\end{equation}
where $\mathbf{R}^{\mathrm{conv}}$, $\mathbf{R}^{\mathrm{diff}}$, $\mathbf{R}^{\mathrm{pres}}$ are computed via~\eqref{eq:message}--\eqref{eq:aggregate}. This is a standard MP residual layer: embarrassingly parallel over all edges. The resulting $\mathbf{u}^*$ violates continuity in general.

\paragraph{Layer 2: Graph Laplacian projection layer}
To restore $\nabla\cdot\mathbf{u}=0$, assembling the implicit discretization of equation~\eqref{eq:fvm_momentum} yields a linear system whose diagonal coefficient $a_i^P$ arises from the time-derivative and implicit diffusion terms, and whose off-diagonal operator $H_i(\mathbf{u})$ collects all neighbor contributions:
\begin{equation}
a_i^P\,\mathbf{u}_i = H_i(\mathbf{u}^*) - \frac{1}{\rho}\sum_{f \in \mathcal{F}(i)} p_f^{n+1}\,\mathbf{S}_f. \label{eq:Ap_H}
\end{equation}
Substituting into the discrete continuity constraint $\sum_{f \in \mathcal{F}(i)}\mathbf{u}_f^{n+1}\cdot\mathbf{S}_f = 0$ and applying Rhie--Chow interpolation for the face velocity yields:
\begin{equation}
\underbrace{\sum_{f \in \mathcal{F}(i)} \frac{1}{a_f^P}(\nabla p^{n+1})_f \cdot \mathbf{S}_f}_{\text{weighted graph Laplacian on }\mathcal{G}} = \sum_{f \in \mathcal{F}(i)} \left(\frac{H(\mathbf{u}^*)}{a^P}\right)_{\!\!f}\!\!\cdot\,\mathbf{S}_f. \label{eq:pressure_poisson}
\end{equation}
The left-hand side is precisely a \emph{weighted graph Laplacian} on $\mathcal{G}$ with edge weights $1/a_f^P$, assembled from the same \texttt{owner}/\texttt{neighbor} arrays via scatter/gather. The right-hand side is a standard MP aggregation of the face-interpolated $H/a^P$ field. The resulting symmetric positive-definite linear system is solved iteratively via conjugate gradient with ILU preconditioning.

\paragraph{Layer 3: Correction MP layer}
Once $p^{n+1}$ is obtained, the velocity field is corrected by a final scatter operation:
\begin{equation}
\mathbf{u}^{n+1}_i = \frac{H_i(\mathbf{u}^*)}{a_i^P} - \frac{1}{\rho\,a_i^P}\sum_{f \in \mathcal{F}(i)} p^{n+1}_f\,\mathbf{S}_f, \label{eq:velocity_correction}
\end{equation}
and the face-normal flux is corrected via Rhie--Chow interpolation to enforce discrete mass conservation at each face. Layers~2--3 can be repeated (typically twice) to reduce the pressure--velocity splitting error. The complete three-layer procedure constitutes $\mathcal{U}$ in equation~\eqref{eq:node_update}: a divergence-free constrained update implemented entirely as graph operations on $\mathcal{G}$.

\subsection{End-to-End Automatic Differentiation}
\label{sec:ad}

\subsubsection{Reverse-mode differentiation through the solver}

End-to-end gradient computation in DiFVM is provided by JAX's reverse-mode (vector-Jacobian product, VJP) automatic differentiation, applied to the complete simulation pipeline. Given a scalar loss function $\mathcal{L}(\mathbf{u}(T;\boldsymbol{\theta}))$ that depends on the simulated state at time $T$ through parameters $\boldsymbol{\theta}$ (which may include boundary condition values, physical parameters, model coefficients, or initial conditions), the gradient $\partial\mathcal{L}/\partial\boldsymbol{\theta}$ is computed via \texttt{jax.vjp} for the simulation function:
\begin{equation}
\frac{\partial \mathcal{L}}{\partial \boldsymbol{\theta}} = \mathbf{v}^{\top} \frac{\partial \mathbf{u}(T;\boldsymbol{\theta})}{\partial \boldsymbol{\theta}}, \label{eq:vjp}
\end{equation}
where $\mathbf{v} = \partial\mathcal{L}/\partial\mathbf{u}(T)$ is the adjoint vector (loss sensitivity with respect to the final state). JAX's AD engine propagates this adjoint backward through the \texttt{jax.lax.scan} loop and through every graph-based operator, including convective fluxes, diffusive fluxes, pressure solves, velocity corrections, and boundary condition applications, without any manual adjoint derivation. Because all operations are expressed as differentiable JAX primitives (arithmetic, gather, scatter, linear solves), the VJP rules are automatically composed to yield exact machine-precision gradients.

\subsubsection{Gradient checkpointing and implicit differentiation}

Naive automatic differentiation through an $n_\mathrm{steps}$-step time-stepping  requires storing all intermediate activations, incurring $\mathcal{O}(n_\mathrm{steps})$ memory overhead. For long simulations on fine meshes, this can exceed GPU memory capacity. DiFVM addresses this via gradient checkpointing, which selectively rematerializes intermediate states during the backward pass rather than storing them all:
\begin{equation}
\text{Memory} \sim \mathcal{O}(\sqrt{n_\mathrm{steps}} \cdot N_c), \qquad \text{Compute} \sim \mathcal{O}(n_\mathrm{steps} \cdot N_c), \label{eq:checkpointing}
\end{equation}
using the optimal binomial checkpointing schedule~\cite{blondel2024elements} applied within the \texttt{jax.lax.scan} loop. At each PISO inner iteration, implicit differentiation is additionally applied through the pressure Poisson linear solve to avoid unrolling the iterative solver (which would require storing all conjugate-gradient iterates). Specifically, for a converged pressure solve $\mathcal{F}(p^*,\mathbf{u}^*) = 0$, the adjoint of $p^*$ with respect to $\mathbf{u}^*$ is computed via the implicit function theorem~\cite{akhare2025implicit}:
\begin{equation}
\frac{\partial p^*}{\partial \mathbf{u}^*} = -\left(\frac{\partial \mathcal{F}}{\partial p^*}\right)^{-\!1} \frac{\partial \mathcal{F}}{\partial \mathbf{u}^*}, \label{eq:implicit_diff}
\end{equation}
where the adjoint linear system $(\partial\mathcal{F}/\partial p^*)^\top\boldsymbol{\lambda} = \partial\mathcal{L}/\partial p^*$ is solved using the same iterative solver as the forward pass. This yields exact pressure-solve gradients with memory and compute cost equivalent to a single forward solve, independent of the number of linear solver iterations.

\subsection{Supported Boundary Conditions, Numerical Schemes, and Interfaces}
\label{sec:schemes}

\subsubsection{Numerical schemes and linear solvers}
DiFVM supports a comprehensive set of numerical schemes for time integration, convection discretization, and diffusion flux treatment, summarized in Table~\ref{tab:schemes}. For time integration, first-order forward and backward Euler schemes and the second-order Crank--Nicolson scheme are available; implicit schemes are recommended for stability on fine meshes. Convection schemes range from first-order upwind to higher-order formulations (SOU, QUICK), all implemented within the deferred-correction framework of equation~\eqref{eq:deferred_correction} to preserve diagonal dominance of the momentum matrix. Diffusive flux treatment supports non-orthogonality correction strategies as described in Section~\ref{sec:fvm-flux}: no correction (orthogonal meshes only), minimum correction, orthogonal correction, and over-relaxed correction adopted by default. Pressure--velocity coupling is handled exclusively by the PISO algorithm. The pressure Poisson and momentum linear systems are solved by conjugate gradient (CG), BiCGStab, or GMRES, each with ILU preconditioning; CG is the default for the symmetric pressure system and BiCGStab for the non-symmetric momentum system.

\begin{table}[htb!]
\centering
\caption{Supported numerical models, discretization schemes, and features in DiFVM.}
\label{tab:schemes}
\footnotesize
\begin{tabular}{ll}
\toprule
\textbf{Category} & \textbf{Supported options} \\
\midrule
Mesh & Unstructured polyhedral, arbitrary element types \\
Time integration & Forward Euler (1st), Backward Euler (1st implicit), Crank--Nicolson (2nd) \\
Pressure--velocity & PISO \\
Convection & Upwind (1st), Central (2nd), SOU (deferred correction), QUICK (deferred correction) \\
Diffusion flux & No correction, minimum correction, orthogonal correction, over-relaxed (default) \\
Linear solvers & CG, BiCGStab, GMRES (all with ILU preconditioning) \\
Boundary conditions & Dirichlet, Neumann, empty (2D), time-varying inflow, Windkessel RCR \\
Interface & OpenFOAM drop-in (\texttt{blockMesh}/\texttt{snappyHexMesh} compatible), VTK unstructured grids \\
\bottomrule
\end{tabular}
\end{table}

\subsubsection{Standard boundary conditions}
\label{sec:bc}
DiFVM supports Dirichlet velocity conditions (fixed-value inlet, no-slip wall), Neumann pressure conditions (zero-gradient outlet, free-slip wall), empty patches for 2D simulations, and time-varying inflow profiles. All boundary conditions are implemented as graph operations on boundary edge patches within the same scatter framework as interior flux computation, and are fully differentiable with respect to their parameters through the VJP pipeline described in Section~\ref{sec:ad}.

\subsubsection{Windkessel RCR outlet condition for cardiovascular applications}
\label{sec:windkessel}
For cardiovascular simulations, DiFVM implements the three-element Windkessel (RCR)  model at outlet boundaries to represent the lumped hemodynamic resistance and compliance of downstream vasculature not explicitly resolved in the computational domain. The three elements represent the proximal arterial resistance $R_p$, the elastic storage capacity of the downstream arterial system $C$, and the peripheral 
vascular resistance $R_d$ of smaller arteries and capillaries. The outlet flow rate  through patch $\Gamma_o$ is computed from the discrete face fluxes:
\begin{equation}
Q_o(t) = \sum_{f \in \Gamma_o} \dot{m}_f, \qquad \dot{m}_f = \mathbf{u}_f \cdot \mathbf{S}_f,
\label{eq:windkessel_Q}
\end{equation}
using the same scatter operation as interior flux aggregation. Let $p_o(t)$ be the area-averaged outlet pressure. The RCR model introduces an internal capacitor pressure state $p_c(t)$ governed by:
\begin{subequations}
\label{eq:windkessel_ode}
\begin{align}
C\frac{dp_c}{dt} + \frac{1}{R_d}p_c = Q_o(t), \label{eq:windkessel_ode_a} \\
p_o(t) = p_c(t) + R_p Q_o(t). \label{eq:windkessel_ode_b}
\end{align}
\end{subequations}
where equation~\eqref{eq:windkessel_ode_a} enforces the flow split between the compliant element and distal resistance, and equation~\eqref{eq:windkessel_ode_b} adds the proximal resistive pressure drop. Since equation~\eqref{eq:windkessel_ode_a} is a linear ODE with constant coefficients, assuming $Q_o$ is piecewise constant over $[t^n, t^{n+1}]$ yields an exact update:
\begin{equation}
p_c^{n+1} = p_c^n\exp\!\left(-\frac{\Delta t}{R_d C}\right) 
+ R_d Q_o^{n+1}\!\left(1 - \exp\!\left(-\frac{\Delta t}{R_d C}\right)\right),
\label{eq:windkessel_discrete}
\end{equation}
and $p_o^{n+1} = p_c^{n+1} + R_p Q_o^{n+1}$ is imposed as the outlet pressure boundary condition in the PISO pressure solve. DiFVM also supports explicit Forward Euler and implicit Backward Euler time discretizations for the Windkessel ODE, which can be selected through the solver configuration. For multiple outlet patches $\{\Gamma_{o,j}\}_{j=1}^{N_o}$, an independent RCR model $(R_{p,j}, C_j, R_{d,j})$ is assigned to each outlet and updated in parallel.

Critically, all parameters $(R_p, C, R_d)$ are exposed as leaf nodes in the JAX computational graph. Since equation~\eqref{eq:windkessel_discrete} is implemented entirely in differentiable JAX operations and $Q_o^{n+1}$ is a differentiable function of the velocity state, the gradient $\partial\mathcal{L}/\partial(R_p, C, R_d)$ is computed automatically by the same VJP pass described in Section~\ref{sec:ad}, enabling simultaneous end-to-end inference of Windkessel parameters across all outlets from sparse clinical observations.

\subsubsection{OpenFOAM and VTK interfaces}
\label{sec:openfoam}
DiFVM provides two levels of external format support. For OpenFOAM, DiFVM operates as a drop-in replacement at the case level: it directly reads a standard OpenFOAM case directory without any modification, ingesting the mesh topology from \texttt{constant/polyMesh/} (\texttt{points}, \texttt{faces}, \texttt{owner}, 
\texttt{neighbour}, \texttt{boundary}), initial and boundary conditions from \texttt{0/U} and \texttt{0/p}, physical parameters from \texttt{constant/transportProperties}, and numerical settings from \texttt{controlDict, fvSchemes, fvSolution}. Users can therefore construct and configure a simulation entirely within the OpenFOAM ecosystem, using \texttt{blockMesh}, \texttt{snappyHexMesh}, or any third-party mesher that exports to OpenFOAM format, and pass the resulting case directory directly to DiFVM without re-specifying mesh, boundary conditions, or physical/numerical parameters. This ensures that cross-solver validation is conducted under rigorously identical problem settings, removing mesh discretization and boundary condition specification as confounding variables in accuracy comparisons. For general unstructured meshes outside the OpenFOAM ecosystem, DiFVM additionally accepts and exports the VTK unstructured grid format, enabling interoperability with third-party mesh generators and pre- and post-processing pipelines. We also provide an interface for SimVascular meshes, allowing seamless integration of SimVascular’s cardiovascular meshing workflow with the DiFVM solver. In all cases, the ingested mesh topology is converted into the same static graph arrays (see Section~\ref{sec:isomorphism}), with no format-specific modifications to the solver core.

\section{Numerical Results}
\label{sec:results}

In this section, we evaluate the accuracy, robustness, and differentiability of DiFVM through a series of benchmark problems organized into two parts. Part~I (Section~\ref{sec:forward}) validates forward simulation accuracy across six test cases of increasing geometric and physical complexity, from a three-dimensional Poisson problem and two-dimensional canonical flows to unsteady cylinder vortex shedding and three-dimensional pulsatile patient-specific aortic hemodynamics. In all cases, DiFVM results are compared quantitatively against OpenFOAM under identical mesh, boundary condition, and numerical settings, with errors evaluated against either analytical solutions or high-fidelity reference data. Part~II (Section~\ref{sec:inverse}) demonstrates end-to-end differentiability through two inverse problems of increasing complexity: recovery of a single scalar boundary parameter from sparse interior velocity measurements in cavity flow, and simultaneous inference of six Windkessel parameters from inlet pressure and outlet flow-rate observations in a pulsatile bifurcation geometry. Together, these benchmarks establish DiFVM as both a high-fidelity GPU-accelerated forward solver on unstructured meshes and a gradient-capable platform for PDE-constrained inverse problems and data-driven hybrid CFD-ML modeling.

\subsection{Forward Simulation Validation}
\label{sec:forward}

\subsubsection{Three-dimensional Poisson equation}
\label{sec:case_poisson}
We first validate DiFVM using a three-dimensional Poisson problem with a prescribed source term and Dirichlet boundary conditions. The computational domain is the unit cube $\Omega = [0,1]\times[0,1]\times[0,1]$, discretized using an unstructured tetrahedral mesh. The governing equation is
\begin{equation}
\nabla \cdot (\nabla \phi) = f \quad \text{in } \Omega,
\label{eq:poisson_3d}
\end{equation}
with Dirichlet boundary conditions imposed on all six faces,
\begin{equation}
\phi = 10 \quad \text{on } \partial \Omega.
\label{eq:poisson_bc}
\end{equation}
The volumetric source term $f$ is prescribed analytically as
\begin{equation}
\begin{aligned}
f(x,y,z) &= -10\pi^2 \sin(\pi x)\,(y^4 - y)\sin(2\pi z) + 24\,\sin(\pi x)\,y^2\,\sin(2\pi z),
\end{aligned}
\label{eq:poisson_source}
\end{equation}
%
which corresponds to the exact solution
\begin{equation}
\phi_{\text{GT}}(x,y,z) = 2\sin(\pi x)\,(y^4 - y)\,\sin(2\pi z) + 10.
\label{eq:poisson_gt}
\end{equation}
By construction, $\phi_{\text{GT}}$ satisfies both equation~\eqref{eq:poisson_3d} and the boundary conditions~\eqref{eq:poisson_bc} exactly, and therefore serves as the analytical ground truth for verification.

The Poisson equation is solved using DiFVM with a second-order accurate finite-volume discretization of the diffusion operator, where surface-normal gradients are computed using a central-difference scheme with over-relaxed non-orthogonal correction. The numerical solution is compared against $\phi_{\text{GT}}$ to evaluate pointwise error distributions and global error norms.
\begin{figure}[htb!]
\centering
\includegraphics[width=\linewidth]{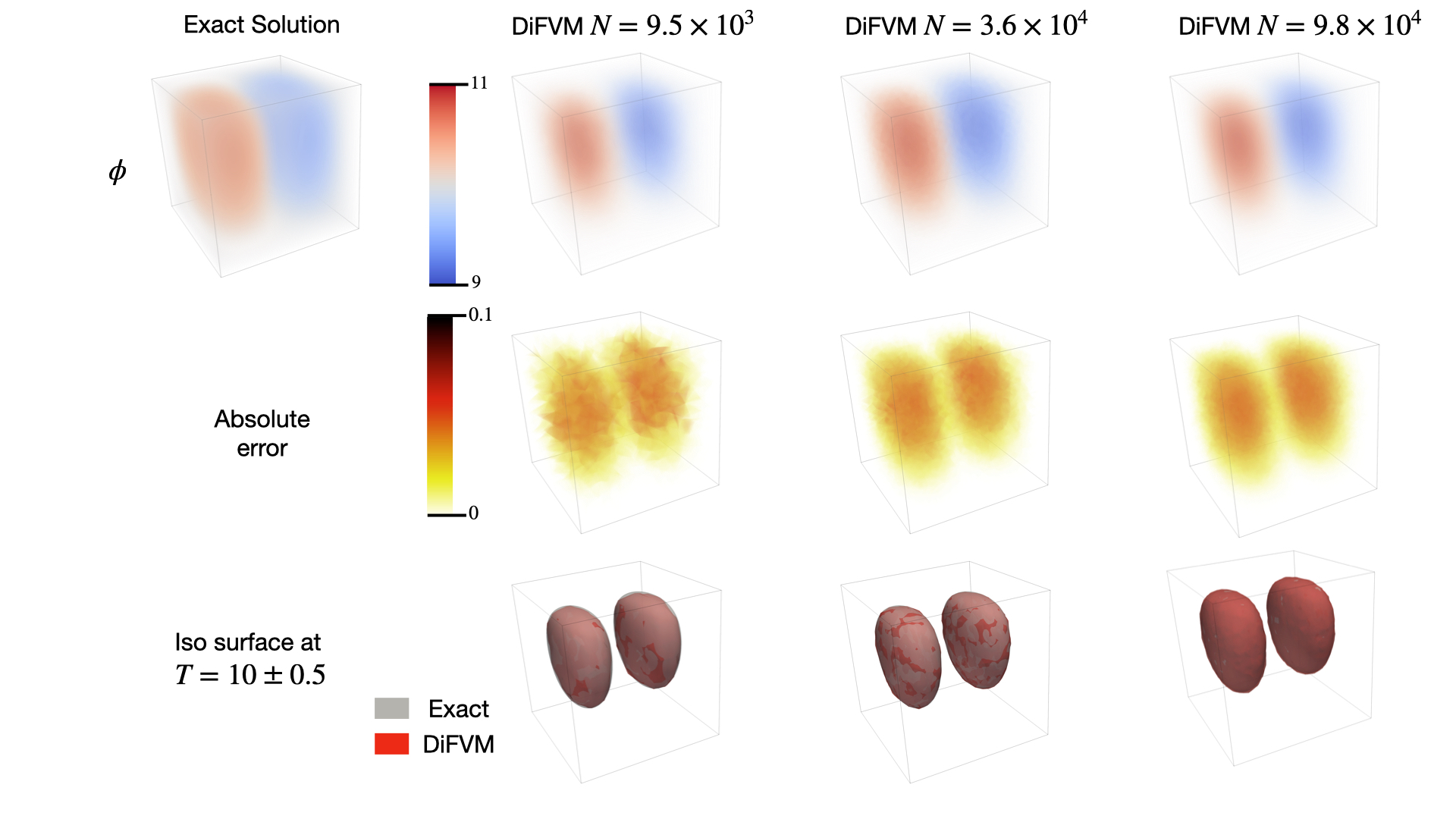}
\caption{Three-dimensional Poisson benchmark. Comparison between the analytical solution and DiFVM results at increasing mesh resolutions ($N = 9.5\times10^3$, $3.6\times10^4$, $9.8\times10^4$ cells, left to right). Top row: scalar field $\phi$. Middle row: absolute error distribution. Bottom row: isosurfaces at $\phi = 10\pm0.5$ (gray: exact; red: DiFVM). Mesh refinement yields monotonically reduced error magnitude and improved isosurface alignment, demonstrating second-order spatial accuracy on unstructured meshes.}
\label{fig:case1}
\end{figure}
Figure~\ref{fig:case1} presents a qualitative comparison between the exact solution and DiFVM results at increasing mesh resolutions ($N = 9.5\times10^3$, $3.6\times10^4$, $9.8\times10^4$ cells). As the grid is refined, the DiFVM solution progressively converges toward the analytical solution, exhibiting improved agreement in the scalar field distribution (top row), reduced absolute error magnitude (middle row), and enhanced alignment of isosurfaces at $\phi = 10 \pm 0.5$ (bottom row). The absolute error decreases monotonically with mesh refinement, confirming the expected convergence behavior of the second-order spatial discretization. Residual errors are primarily concentrated near regions of strong curvature and large gradients, particularly around the extrema associated with the two elliptical core structures along the $z$-direction. This behavior is consistent with amplification of truncation error in regions of high second-derivative magnitude, which diminishes as mesh resolution improves. The results verify the diffusion discretization in DiFVM and demonstrate grid convergence on unstructured meshes.

\subsubsection{Two-dimensional passive scalar advection}
\label{sec:case_advection}
\begin{figure}[htb!]
\centering
\includegraphics[width=\linewidth]{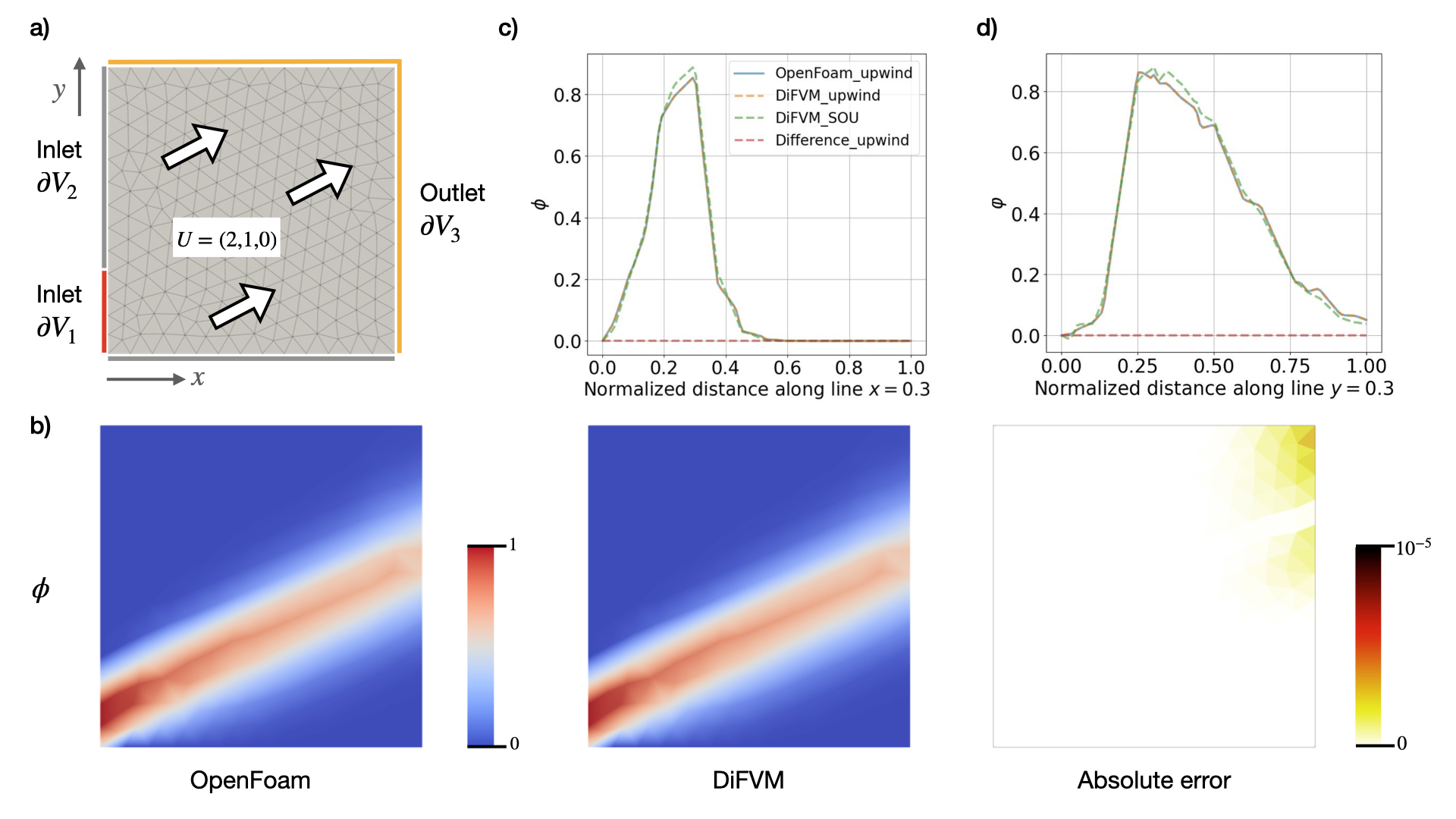}
\caption{Step advection benchmark: (a) computational domain and boundary conditions; (b) scalar field $\phi$ computed by OpenFOAM and DiFVM, together with the absolute error distribution; (c--d) scalar profiles along $x=0.3$ and $y=0.3$, respectively, comparing upwind and SOU schemes.}
\label{fig:step_profile}
\end{figure}

We consider a two-dimensional steady advection--diffusion problem for a passive scalar. The computational domain is a unit square $\Omega=[0,1]\times[0,1]$, discretized using an unstructured triangular mesh, as illustrated in Fig.~\ref{fig:step_profile}(a). A step profile is prescribed at the inlet boundary, with $\phi=1$ on the lower inlet segment ($\partial V_1$) and $\phi=0$ on the upper inlet segment ($\partial V_2$). Homogeneous Neumann boundary conditions are imposed at the outlet ($\partial V_3$). The numerical solution obtained with DiFVM is validated against a reference solution computed using OpenFOAM's \texttt{scalarTransportFoam} solver under identical numerical settings. The governing equation for the passive scalar $\phi$ is
\begin{equation}
\frac{\partial \phi}{\partial t} + \nabla \cdot (\mathbf{u}\phi) - \nabla \cdot (\Gamma \nabla \phi) = f,
\end{equation}
where $\mathbf{u}=(2,1,0)$ is a constant oblique velocity field and $\Gamma = 0.001$ is the diffusion coefficient. These parameters correspond to a P\'{e}clet number $\mathrm{Pe}\approx 2236$, indicating a strongly convection-dominated transport regime. Both DiFVM and OpenFOAM employ a backward Euler time discretization, Gauss linear gradient and interpolation schemes, and over-relaxed corrected surface-normal gradients for the diffusion operator. For the convection term, results are obtained using both the first-order upwind scheme and the SOU scheme implemented via deferred correction.

Figure~\ref{fig:step_profile}(b) compares the steady-state scalar field $\phi$ computed by DiFVM and OpenFOAM, together with the corresponding absolute error distribution. DiFVM accurately captures the oblique advection and diffusion of the inlet step profile, closely matching the OpenFOAM reference solution. The absolute error remains below $10^{-5}$ throughout the domain, confirming numerical equivalence between the two solvers. Figures~\ref{fig:step_profile}(c) and (d) show scalar profiles extracted along the vertical line $x=0.3$ and the horizontal line $y=0.3$, respectively. When using the upwind scheme, DiFVM and OpenFOAM produce nearly identical profiles, with error magnitude below $1\times10^{-5}$. The observed residual differences are attributed to minor numerical perturbations arising from differences in floating-point initialization, subsequently amplified by variations in linear solver backends. In contrast, the SOU scheme yields higher peak values and a sharper transition across the step, indicating reduced numerical diffusion and improved resolution of steep gradients. Overall, this benchmark validates the convective discretization of DiFVM in high-P\'{e}clet-number regimes.

\subsubsection{Two-dimensional elbow flow}
\label{sec:case_elbow}
\begin{figure}[htb!]
\centering
\includegraphics[width=\linewidth]{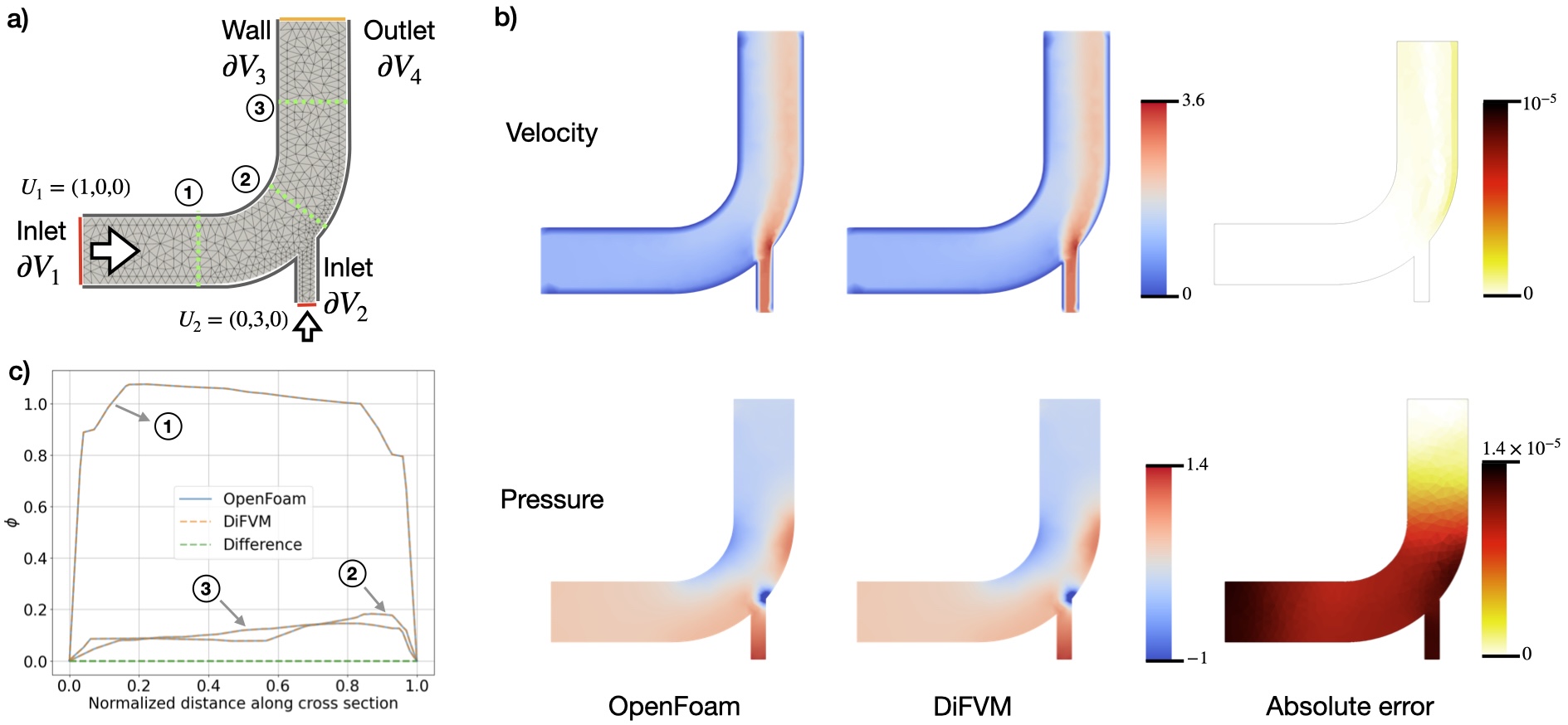}
\caption{Benchmark validation of steady incompressible flow through a two-dimensional elbow geometry. (a) Computational domain and boundary conditions, including prescribed inlet velocities and outlet pressure condition. (b) Comparison of steady-state velocity magnitude (top row) and pressure (bottom row) fields obtained using OpenFOAM and DiFVM, together with the corresponding absolute error distribution. (c) Velocity profiles extracted along representative cross-sections indicated in panel~(a), showing close agreement between DiFVM and OpenFOAM.}
\label{fig:elbow}
\end{figure}

We next consider a two-dimensional incompressible flow through a $90^{\circ}$ elbow, a classical benchmark used to assess the robustness of pressure--velocity coupling, convection discretization, and diffusion accuracy in curved geometries. The computational domain and boundary configuration are illustrated in Fig.~\ref{fig:elbow}(a). The domain is discretized using an unstructured triangular mesh, with no-slip boundary conditions applied on the walls ($\partial V_3$). Two velocity inlets are prescribed at the lower horizontal ($\partial V_1$) and vertical ($\partial V_2$) branches, with uniform inflow velocities $\mathbf{u}_1=(1,0,0)$ and $\mathbf{u}_2=(0,3,0)$, respectively. A homogeneous Neumann boundary condition is imposed at the outlet ($\partial V_4$) for pressure. The governing equations are the incompressible Navier--Stokes equations~\eqref{eq:continuity}--\eqref{eq:momentum}, solved using a backward Euler time discretization. Both DiFVM and OpenFOAM employ identical numerical settings, including Gauss linear gradient and interpolation schemes, upwind discretization for the convective term, over-relaxed corrected surface-normal gradients for diffusion, and PISO pressure--velocity coupling with Rhie--Chow flux correction. The velocity and pressure linear systems are solved using PCG and PBiCGStab solvers, respectively.

Figure~\ref{fig:elbow}(b) compares the steady-state velocity magnitude and pressure fields obtained using DiFVM and OpenFOAM, together with the corresponding absolute error distributions. DiFVM accurately reproduces the characteristic acceleration and redistribution of velocity through the curved section of the elbow, as well as the pressure buildup along the outer wall of the bend. The absolute error remains close to machine precision throughout the domain. To further quantify the agreement, Fig.~\ref{fig:elbow}(c) presents velocity profiles extracted along three representative cross-sections. Across all sections, DiFVM predictions closely match the OpenFOAM reference solution, capturing both the core flow behavior and near-wall velocity gradients. Since DiFVM and OpenFOAM follow the same mathematical formulation and PISO pressure--velocity coupling procedure, the remaining discrepancies arise primarily from differences in floating-point precision at early computational stages, subsequently amplified by variations in linear solvers and underlying algebraic operations. These effects remain negligible, with errors consistently below $1\times10^{-5}$. Overall, this benchmark validates the robustness and accuracy of the pressure--velocity coupling in DiFVM on irregular geometries.

\subsubsection{Two-dimensional lid-driven cavity flow}
\label{sec:case_cavity}

\begin{figure}[htb!]
\centering
\includegraphics[width=\linewidth]{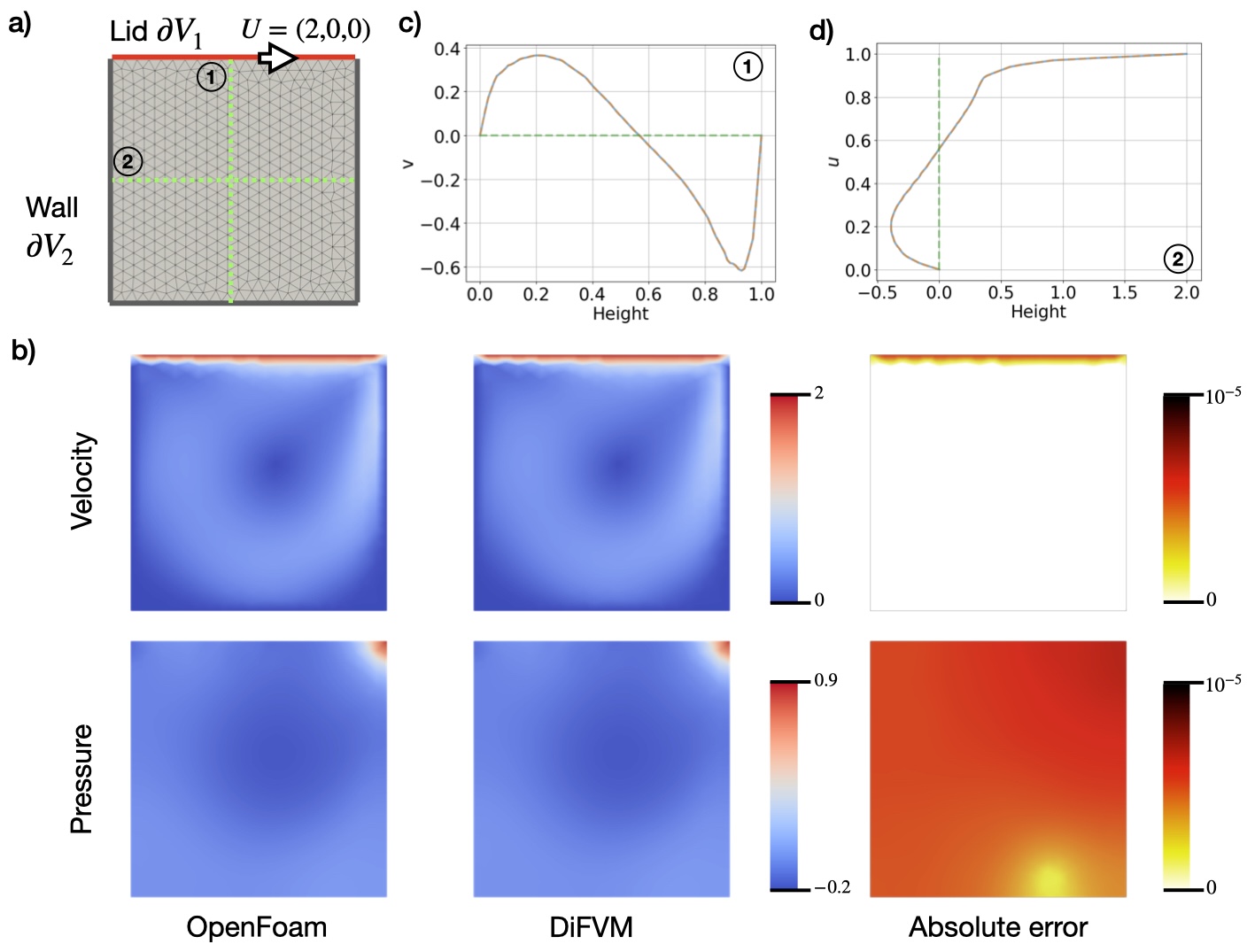}
\caption{Benchmark validation of the two-dimensional lid-driven cavity flow. (a) Computational domain, unstructured mesh, and boundary conditions, including the moving lid. (b) Comparison of steady-state velocity magnitude and pressure fields obtained using OpenFOAM and DiFVM, together with the corresponding absolute error distributions. (c) Velocity profile along the vertical centerline of the cavity. (d) Velocity profile along the horizontal centerline. Excellent agreement is observed between DiFVM and OpenFOAM in both field distributions and centerline profiles.}
\label{fig:cavity}
\end{figure}

We next consider the classical two-dimensional lid-driven cavity flow, a canonical benchmark characterized by the formation of flow vortices and strong pressure--velocity coupling. The computational domain is a unit square $\Omega = [0,1]\times[0,1]$, discretized using an unstructured triangular mesh, as illustrated in Fig.~\ref{fig:cavity}(a). The flow is driven by a moving lid at the top boundary, where a uniform horizontal velocity $\mathbf{u}=(2,0)$ is prescribed, while all remaining walls satisfy no-slip boundary conditions. The governing incompressible Navier--Stokes equations~\eqref{eq:continuity}--\eqref{eq:momentum} are solved using a backward Euler time discretization and the PISO algorithm for pressure--velocity coupling.

Figure~\ref{fig:cavity}(b) compares the steady-state velocity magnitude and pressure fields obtained with DiFVM against reference solutions computed using OpenFOAM under identical numerical settings. DiFVM accurately captures the primary recirculation vortex induced by the moving lid, as well as the associated pressure distribution, including the low-pressure core near the vortex center and pressure buildup near the cavity corners. The absolute error fields confirm that discrepancies between DiFVM and OpenFOAM remain small. Figures~\ref{fig:cavity}(c) and (d) present velocity profiles extracted along the vertical and horizontal centerlines of the cavity, respectively. DiFVM closely matches the OpenFOAM results for both velocity components, accurately reproducing the expected extrema, inflection points, and sign changes associated with the primary vortex structure. This benchmark validates the accuracy of DiFVM in the presence of vortex-dominated flows.

\subsubsection{Two-dimensional flow past a cylinder}
\label{sec:case_cylinder}

We next consider the classical two-dimensional incompressible flow past a circular cylinder at Reynolds number $Re=200$, a benchmark that exhibits unsteady vortex shedding and provides a stringent validation of the transient Navier--Stokes solver. The computational domain consists of a circular cylinder of diameter $D = 0.2$ embedded in a rectangular channel. A uniform inflow velocity $U_\infty = 1$ is prescribed at the inlet, no-slip boundary conditions are imposed on the cylinder surface, and a zero-gradient outlet condition is applied at the outflow. The unsteady incompressible Navier--Stokes equations~\eqref{eq:continuity}--\eqref{eq:momentum} are solved using backward Euler time integration and the PISO algorithm for pressure--velocity coupling. Convection is discretized using the upwind scheme with deferred correction, and diffusion is discretized with an over-relaxed corrected surface-normal gradient scheme to account for mesh non-orthogonality. The time step is $\Delta t = 10^{-3}$, and the simulation is advanced for $20{,}000$ time steps to reach a statistically periodic state.

\begin{figure}[htb!]
\centering
\includegraphics[width=0.8\linewidth]{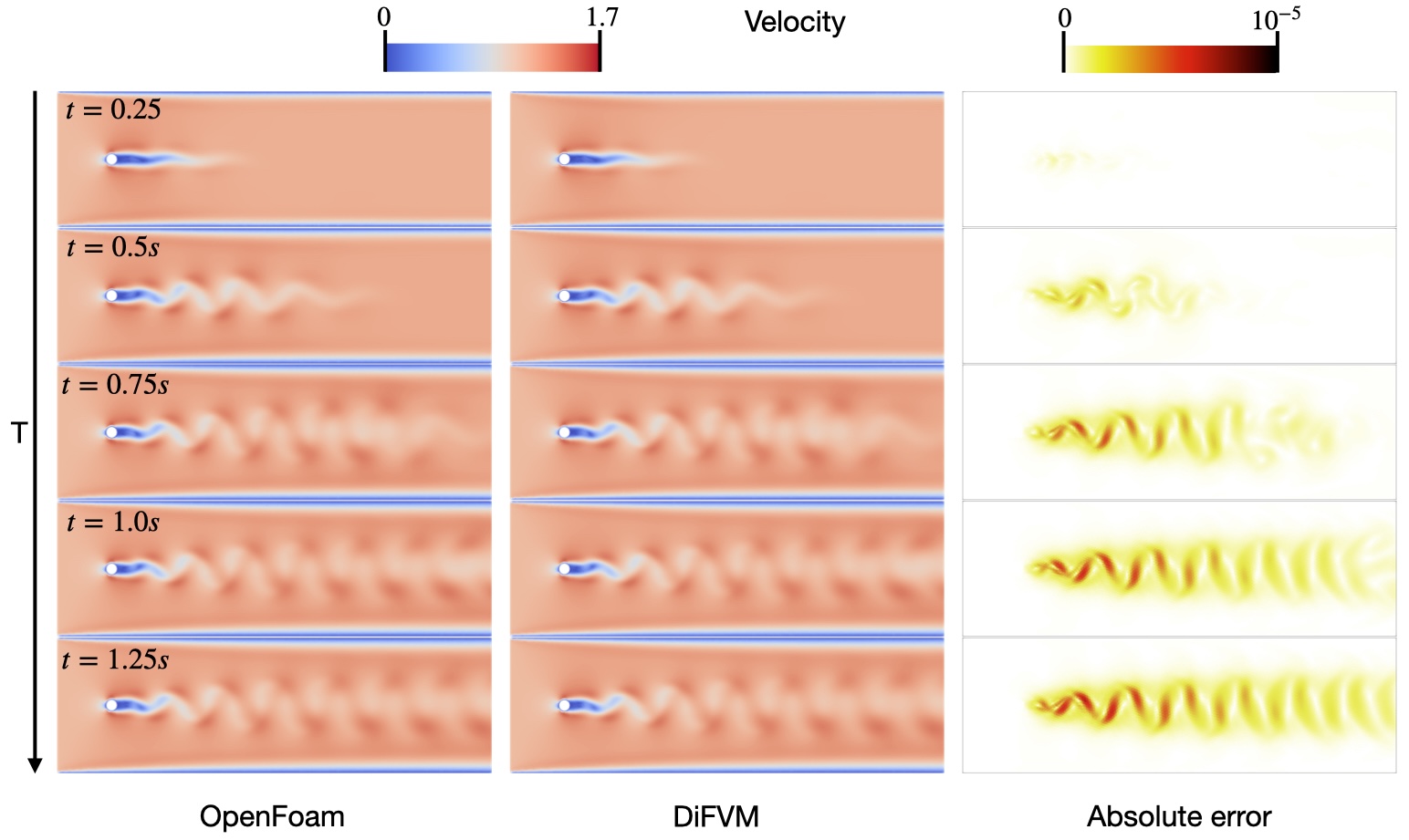}
\caption{Two-dimensional flow past a cylinder at $Re=200$: comparison of instantaneous velocity magnitude fields between OpenFOAM (left) and DiFVM (center) at five successive time instants over one shedding cycle.}
\label{fig:cylinder1}
\end{figure}
\begin{figure}[htb!]
\centering
\includegraphics[width=0.8\linewidth]{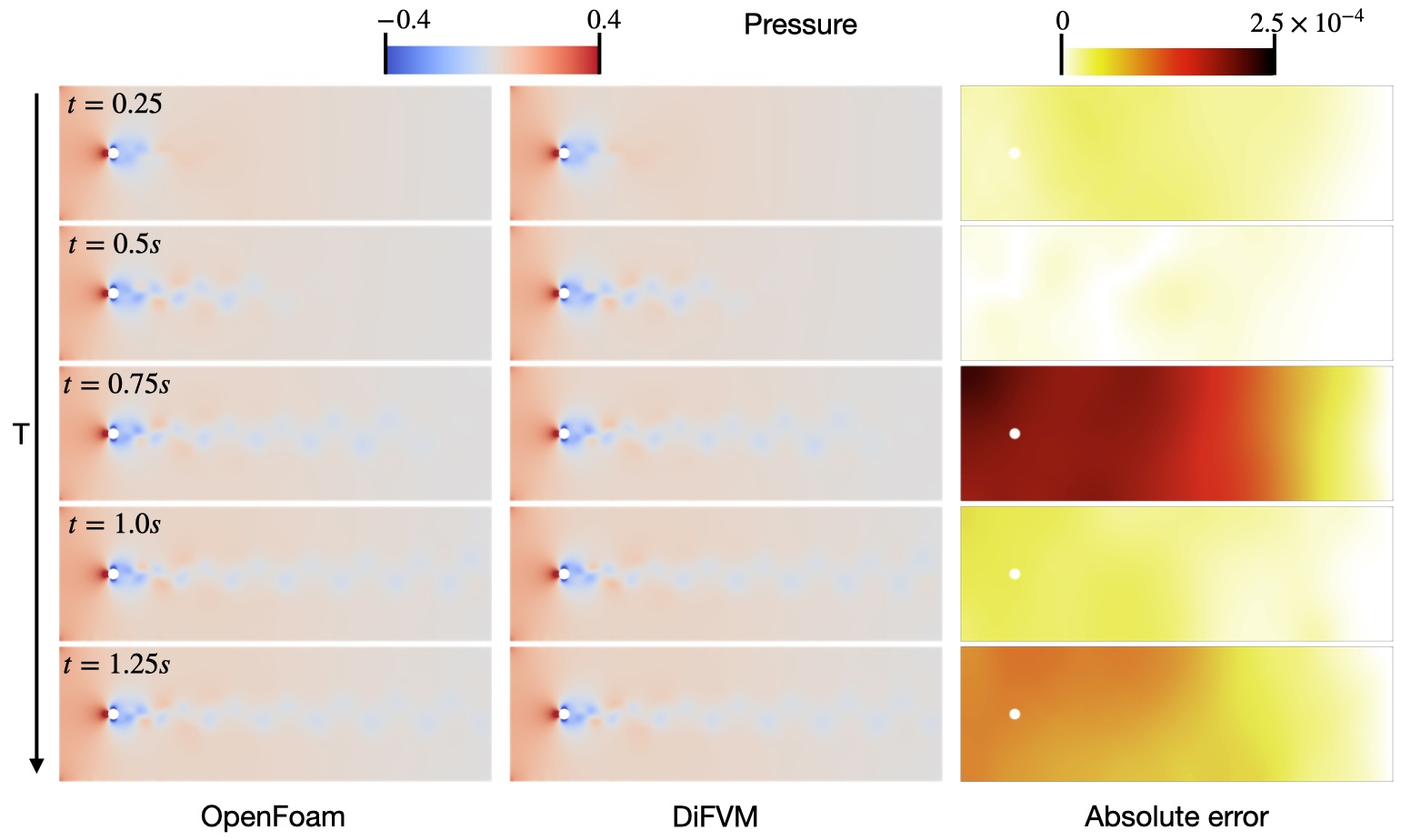}
\caption{Two-dimensional flow past a cylinder at $Re=200$: comparison of instantaneous pressure fields between OpenFOAM (left) and DiFVM (center) at the same time instants as Fig.~\ref{fig:cylinder1}.}
\label{fig:cylinder2}
\end{figure}

Figures~\ref{fig:cylinder1} and \ref{fig:cylinder2} compare instantaneous velocity and pressure fields between DiFVM and OpenFOAM at multiple time instants over one shedding cycle. The results clearly capture the onset, spatial structure, and temporal evolution of periodic vortex shedding in the cylinder wake, consistent with classical observations at $Re=200$. The absolute error in velocity remains below $10^{-5}$ and in pressure below $2.5\times10^{-4}$ throughout the simulation. These results demonstrate that DiFVM accurately resolves unsteady wake dynamics and pressure--velocity coupling in convection-dominated transient flows.

\subsubsection{Three-dimensional patient-specific aortic flow}
\begin{figure}[htb!]
\centering
\includegraphics[width=\linewidth]{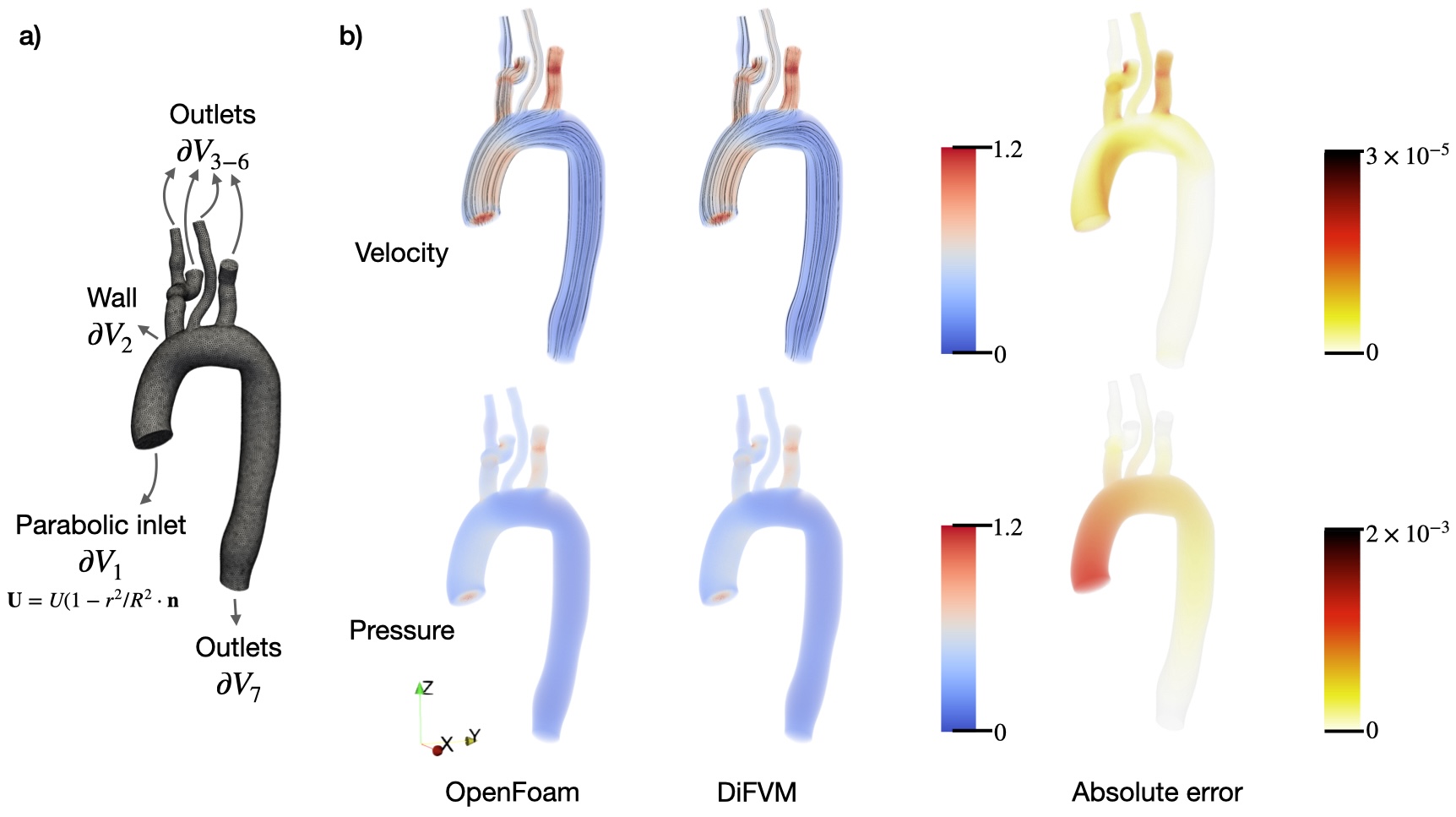}
\caption{Three-dimensional multi-branch aortic flow simulation. (a) Patient-specific aortic geometry with boundary conditions, including a parabolic inlet velocity profile, no-slip vessel walls, and multiple outlet boundaries ($\partial V_{3\text{--}7}$). (b) Comparison of velocity magnitude (top) and pressure (bottom) fields computed using OpenFOAM and DiFVM, together with the absolute error distributions.}
\label{fig:vascular}
\end{figure}

We consider a three-dimensional patient-specific multi-branch aortic geometry obtained from the Vascular Model Repository (VMR)~\cite{Wilson2013}, to demonstrate the capability of DiFVM in handling complex 3D vascular domains. The computational mesh consists of an unstructured tetrahedral discretization of the aortic arch and its major branches, including the right and left common carotid arteries and the right and left subclavian arteries, as illustrated in Fig.~\ref{fig:vascular}(a). A parabolic velocity profile is prescribed at the inlet to represent fully developed laminar inflow,
\begin{equation}
\mathbf{u}(\mathbf{x}) = U_{\max}\!\left(1-\frac{r^2}{R^2}\right)\mathbf{n},
\end{equation}
where $r$ denotes the radial distance from the inlet centerline, $R$ is the inlet radius, and $\mathbf{n}$ is the outward unit normal. No-slip boundary conditions are imposed on the vessel walls. Multiple outlet boundaries are prescribed with zero-gradient (traction-free) conditions to allow natural flow partitioning among the branches. The incompressible Navier--Stokes equations are solved using backward Euler time integration and the PISO algorithm. The DiFVM solution is validated against OpenFOAM under identical numerical schemes and boundary-condition settings.

Figure~\ref{fig:vascular}(b) compares the steady-state velocity and pressure fields obtained using DiFVM and OpenFOAM, together with the corresponding absolute error distributions. DiFVM accurately reproduces the complex three-dimensional flow patterns in the aortic arch, including velocity redistribution induced by curvature and branching, as well as the associated pressure gradients along the vessel walls. The absolute errors remain small and spatially localized ($\leq 3\times10^{-5}$ for velocity, $\leq 2\times10^{-3}$ for pressure), indicating consistent numerical treatment between the two solvers. This benchmark demonstrates that DiFVM robustly handles realistic patient-specific vascular geometries and accurately resolves three-dimensional hemodynamics.

\subsection{Inverse Modeling Demonstrations}
\label{sec:inverse}

\subsubsection{Lid velocity recovery in cavity flow}
\label{sec:case_inv_cavity}

\begin{figure}[htb!]
\centering
\includegraphics[width=\linewidth]{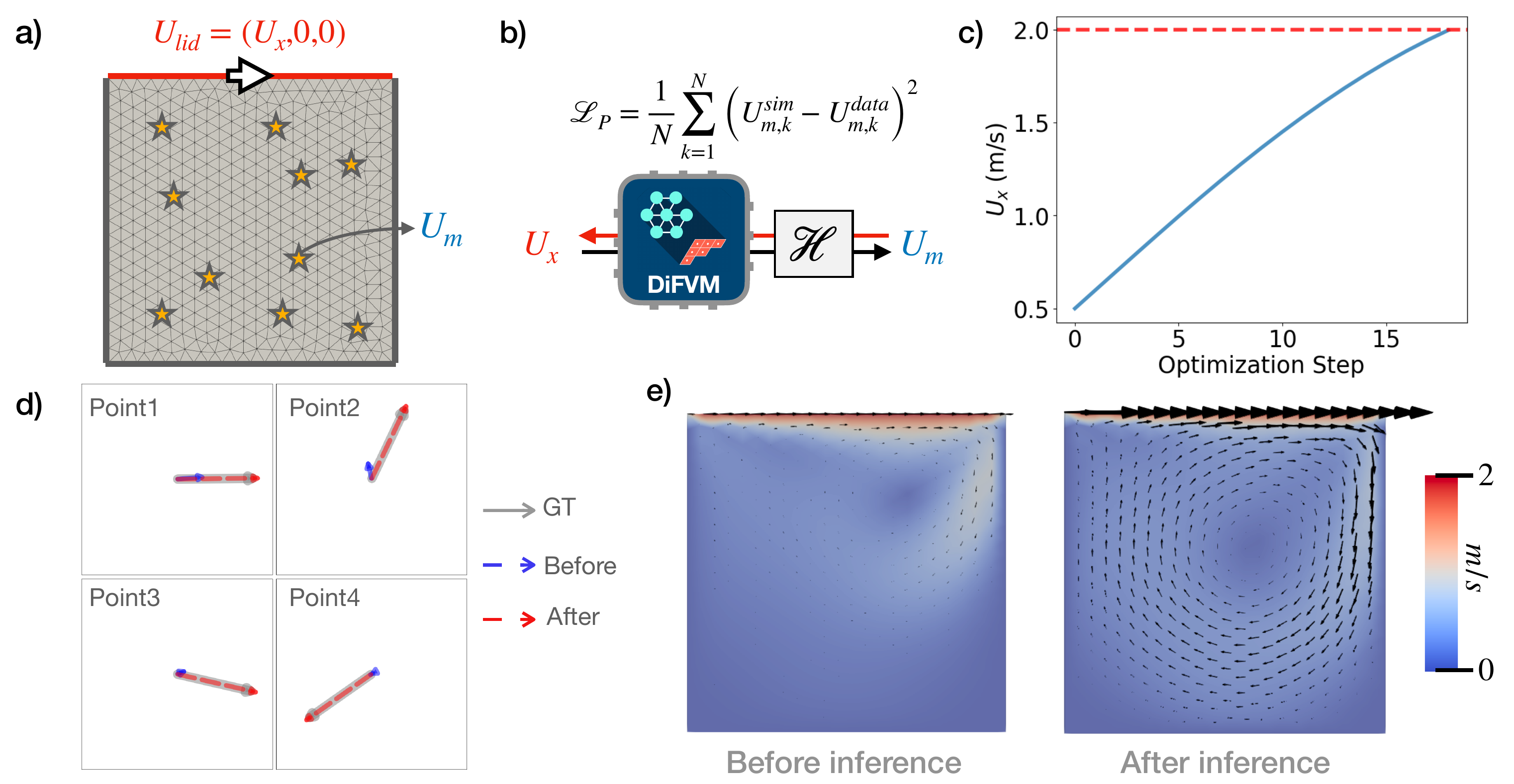}
\caption{Inverse identification of lid velocity using differentiable CFD. (a) Cavity flow configuration with prescribed lid velocity $U_{\mathrm{lid}} = (U_x, 0, 0)$ and interior velocity measurements $U_m$ at 10 randomly selected probe locations. (b) Inverse framework: DiFVM predicts velocities from $U_x$, and the parameter is optimized by minimizing the MSE between simulated and measured velocities. (c) Convergence of the inferred lid velocity toward the ground truth over optimization iterations. (d) Velocity vectors at four representative probe locations before inference (blue), after inference (red), and ground truth (gray). (e) Velocity magnitude contours and vector fields before and after inference.}
\label{fig:inverse_cavity}
\end{figure}

After validating the forward accuracy of DiFVM, we exploit its end-to-end differentiability to solve an inverse problem in the lid-driven cavity flow. As illustrated in Figure~\ref{fig:inverse_cavity}, the objective is to infer the $x$-component of the lid velocity $U_x$ from sparse velocity measurements probed at $N=10$ randomly selected locations within the flow domain. The observation operator is
\begin{equation}
\mathcal{H}(\mathbf{U}) = \mathbf{H}\mathbf{U},
\end{equation}
where $\mathbf{H} \in \mathbb{R}^{N \times N_v}$ is a masking matrix that extracts velocities at the probing locations from the full velocity field $\mathbf{U} \in \mathbb{R}^{N_v \times 2}$, $N_v$ is the total number of mesh cells, and $U_{m,k}$ denotes the $k$-th probed velocity measurement. The loss function is the mean squared error (MSE) between simulated and reference velocities at the probe locations:
\begin{equation}
\mathcal{L} = \frac{1}{N}\sum_{k=1}^{N}\left(U_{m,k}^{\mathrm{sim}} - U_{m,k}^{\mathrm{data}}\right)^2,
\end{equation}
where the synthetic data are generated from a forward simulation with ground truth $U_x = 2.0\,\mathrm{m/s}$.

The ADAM optimizer is employed with an initial learning rate of $1\times10^{-1}$. Starting from an initial guess of $0.5\,\mathrm{m/s}$, the inferred lid velocity converges to $1.996\,\mathrm{m/s}$ within 18 epochs, closely matching the ground truth of $2.0\,\mathrm{m/s}$ (Figure~\ref{fig:inverse_cavity}c). Figure~\ref{fig:inverse_cavity}d shows the initial and optimized velocities at four representative probe locations; after optimization, the inferred velocities align closely with the reference values. Figure~\ref{fig:inverse_cavity}e further confirms that the inverse framework recovers both the global lid velocity and the full flow field with high fidelity.

\subsubsection{Windkessel parameter inference in pulsatile flow}

\begin{figure}[t!]
\centering
\includegraphics[width=\linewidth]{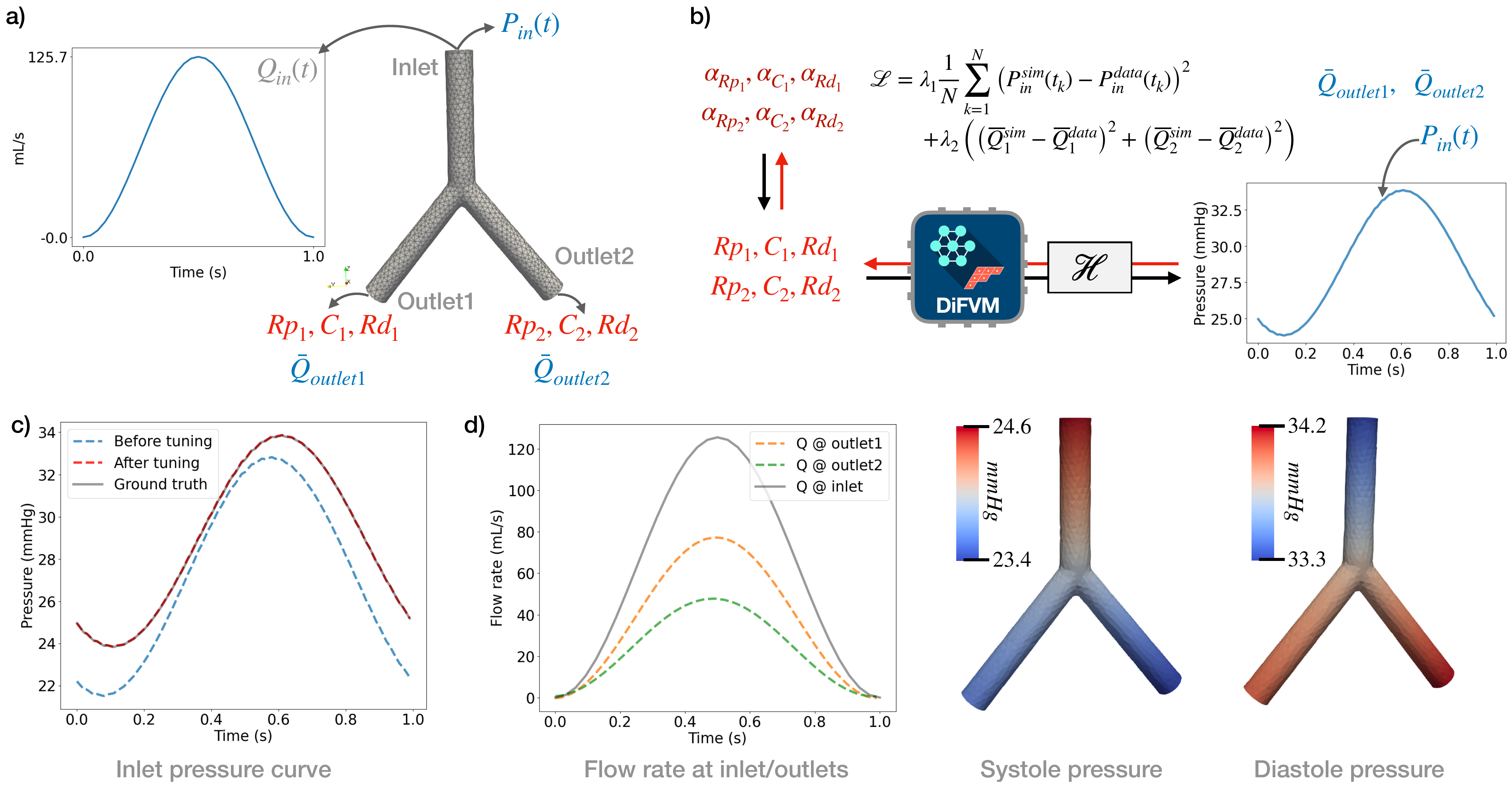}
\caption{Inverse identification of RCR boundary parameters in a bifurcation model. (a) Problem setup: a pulsatile inflow $Q_{\mathrm{in}}(t)$ is prescribed at the inlet, while each outlet is modeled with an RCR boundary condition with parameters $(R_p, C, R_d)$. (b) Inverse framework: outlet parameters $\{R_{p,i}, C_i, R_{d,i}\}$ are optimized by minimizing a composite loss combining the inlet pressure waveform mismatch and time-averaged outlet flow discrepancies. (c) Inlet pressure waveform before tuning (blue), after tuning (red), and ground truth (black), demonstrating accurate recovery of the target pressure curve. (d) Flow rate waveforms at the inlet and both outlets, confirming mass conservation after parameter inference. (e) Pressure fields at systole and diastole after tuning.}
\label{fig:inverse_rcr}
\end{figure}

We further increase the inverse complexity by considering a bifurcation geometry equipped with three-element Windkessel (RCR) boundary conditions at both outlets (Figure~\ref{fig:inverse_rcr}). A pulsatile parabolic velocity profile is prescribed at the inlet. The inverse task is to infer the six Windkessel parameters $\{R_{p1}, C_1, R_{d1}, R_{p2}, C_2, R_{d2}\}$ from the inlet pressure waveform $P(t)$ and the time-averaged flow rates at the two outlets $\{\overline{Q}_1,\overline{Q}_2\}$. 
Traditionally, this task is performed via an empirical estimation~\cite{westerhof2009arterial}. First, the total resistance of each outlet branch is computed from the mean pressure drop and the corresponding outlet flow rate:
\begin{equation}
R_{\mathrm{tot},i} 
= \frac{\overline{P} - P_v}{\overline{Q}_i} 
= \frac{\frac{1}{T}\int_0^T P(t)\,dt - P_v}{\overline{Q}_i}.
\end{equation}

The total resistance is then split into proximal and distal components using a prescribed fraction $\gamma$,
\begin{align}
R_{p,i} &= \gamma R_{\mathrm{tot},i}, \\
R_{d,i} &= (1-\gamma)R_{\mathrm{tot},i},
\end{align}
where $\gamma \in [0.05,0.3]$ is typically used in cardiovascular modeling. 

The compliance is estimated from the exponential decay relation between the Windkessel time constant $\tau$ and the diastolic pressure waveform,
\begin{align}
P(t) - P_v \approx A e^{-t/\tau},
\end{align}
which leads to the compliance
\begin{align}
C_i = \frac{\tau}{R_{d,i}} .
\end{align}

In this study, we adopt $\gamma = 0.15$ for the initial estimation. The resulting parameter values, listed in Table~\ref{tab:rcr_params}, are used as the initial state for the subsequent inverse optimization.

To ensure balanced optimization across parameters of different magnitudes, we optimize logarithmic scaling coefficients $\{\alpha_{R_{p1}}, \alpha_{C_1}, \alpha_{R_{d1}}, \alpha_{R_{p2}}, \alpha_{C_2}, \alpha_{R_{d2}}\}$ defined by
\begin{equation}
R_{pi} = R_{pi}^{\mathrm{gt}}\,e^{\alpha_{R_{pi}}}, \quad C_i = C_i^{\mathrm{gt}}\,e^{\alpha_{C_i}}, \quad R_{di} = R_{di}^{\mathrm{gt}}\,e^{\alpha_{R_{di}}},
\end{equation}
where $i \in \{1,2\}$ denotes the outlet index and the superscript ``gt'' refers to ground truth values. This exponential parametrization ensures strictly positive Windkessel parameters and maps the ground truth to the zero vector, promoting smooth optimization. The observation operator maps the simulated flow field $\{\mathbf{U}, \mathbf{P}\}$ to measurable quantities:
\begin{equation}
\mathcal{H}(\mathbf{U}, \mathbf{P}) = \left(P_{\mathrm{in}}(t),\, \overline{Q}_1,\, \overline{Q}_2\right),
\end{equation}
where the area-averaged inlet pressure is
\begin{equation}
P_{\mathrm{in}}(t) = \frac{\sum_{k \in \partial\Omega_{\mathrm{in}}} P_k(t)\,A_k}{\sum_{k \in \partial\Omega_{\mathrm{in}}} A_k},
\end{equation}
and the time-averaged outlet flow rate at outlet $i$ is
\begin{equation}
\overline{Q}_i = \frac{1}{T}\int_0^T \sum_{k \in \partial\Omega_{\mathrm{out},i}} \mathbf{U}_k(t)\cdot\mathbf{n}_k\,A_k\,dt, \quad i=1,2,
\end{equation}
with $A_k$ and $\mathbf{n}_k$ denoting the area and outward unit normal of the $k$-th surface element. Both pressure and flow quantities are normalized prior to computing the loss to balance gradient magnitudes. The total loss combines the inlet pressure waveform mismatch and the time-averaged outlet flow mismatch:
\begin{equation}
\mathcal{L} = \lambda_1 \mathcal{L}_P + \lambda_2 \mathcal{L}_Q,
\end{equation}
where
\begin{equation}
\mathcal{L}_P = \frac{1}{N}\sum_{k=1}^{N}\left(P_{\mathrm{in}}^{\mathrm{sim}}(t_k) - P_{\mathrm{in}}^{\mathrm{data}}(t_k)\right)^2,
\end{equation}
\begin{equation}
\mathcal{L}_Q = \left(\overline{Q}_1^{\mathrm{sim}} - \overline{Q}_1^{\mathrm{data}}\right)^2 + \left(\overline{Q}_2^{\mathrm{sim}} - \overline{Q}_2^{\mathrm{data}}\right)^2.
\end{equation}

The ADAM optimizer is used with an initial learning rate of $1\times10^{-1}$ and convergence defined as relative error below $1\%$ for all target parameters. The simulation uses $\Delta t = 0.001\,\mathrm{s}$ and total duration $T_{\mathrm{total}} = 6\,\mathrm{s}$ ($N_t = 6000$ time steps), sufficient to reach a stable periodic state. The ground truth and inferred parameters are summarized in Table~\ref{tab:rcr_params}.

\begin{table}[htb!]
\centering
\caption{Ground truth and inferred Windkessel parameters for the two-outlet bifurcation case.}
\label{tab:rcr_params}
\footnotesize
\begin{tabular}{lccccc}
\toprule
\textbf{Parameter} & \textbf{Unit} & \textbf{Initial guess} & \textbf{Ground truth} & \textbf{Inferred} & \textbf{Rel.\ error} \\
\midrule
$R_{p1}$ & dyn$\cdot$s/cm$^5$ & 150.421 & 100.000 & 100.897 & 0.90\% \\
$C_1$ & cm$^5$/(dyn$\cdot$s) & $1.1732\times10^{-3}$ & $1.1111\times10^{-3}$ & $1.1111\times10^{-3}$ & $<0.01\%$ \\
$R_{d1}$ & dyn$\cdot$s/cm$^5$ & 852.386 & 900.000 & 900.000 & $<0.01\%$ \\
$R_{p2}$ & dyn$\cdot$s/cm$^5$ & 240.159 & 160.000 & 160.399 & 0.25\% \\
$C_2$ & cm$^5$/(dyn$\cdot$s) & $7.3481\times10^{-4}$ & $6.9444\times10^{-4}$ & $7.0129\times10^{-4}$ & 0.99\% \\
$R_{d2}$ & dyn$\cdot$s/cm$^5$ & 1360.899 & 1440.000 & 1440.000 & $<0.01\%$ \\
\bottomrule
\end{tabular}
\end{table}

\begin{figure}[htb!]
\centering
\includegraphics[width=\linewidth]{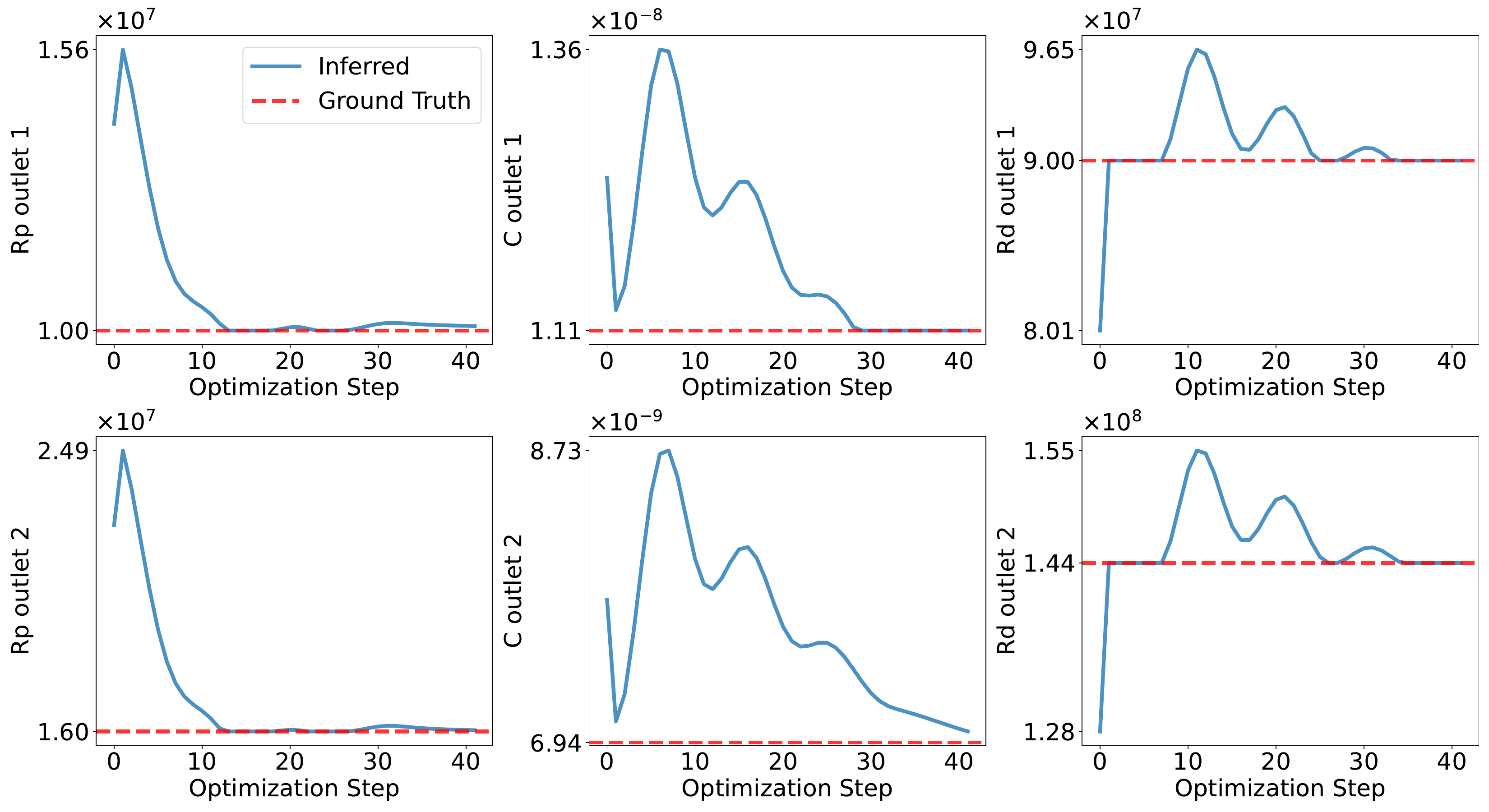}
\caption{Convergence of inferred RCR parameters during optimization. Evolution of proximal resistance $R_p$, compliance $C$, and distal resistance $R_d$ for both outlets over optimization iterations. Solid blue lines denote inferred parameter values; red dashed lines indicate ground truth values. All parameters converge toward their true values within approximately $20$--$30$ iterations, demonstrating stable and accurate Windkessel parameter recovery.}
\label{fig:inverse_rcr_opt}
\end{figure}
All six parameters converge to within $1\%$ of ground truth after 41 optimization epochs (Figure~\ref{fig:inverse_rcr_opt}). Both proximal and distal resistances converge rapidly within approximately 12 epochs, reflecting the strong sensitivity of flow splits to resistance values. Compliance parameters converge more slowly, as the pressure waveform is primarily influenced by capacitance; the distal resistance adjusts transiently to compensate for compliance mismatch before jointly converging with the capacitance parameters. Figure~\ref{fig:inverse_rcr}(c) shows the inlet pressure waveform before and after optimization, demonstrating accurate recovery of the target waveform. Figure~\ref{fig:inverse_rcr}(d) presents the flow rate waveforms at the inlet and both outlets, confirming mass conservation. The pressure contours at systole and diastole (Figure~\ref{fig:inverse_rcr}(e)) illustrate dynamically consistent pressure distributions: during systole, a pronounced pressure drop from inlet to outlets drives forward flow, while during diastole, a temporary reverse pressure gradient develops due to capacitive release in the Windkessel boundary, consistent with pulsatile flow physics.

\section{Discussion}
\label{sec:discussion}

\subsection{Computational Efficiency}
\label{sec:efficiency}

\begin{figure}[htb!]
\centering
\includegraphics[width=0.9\linewidth]{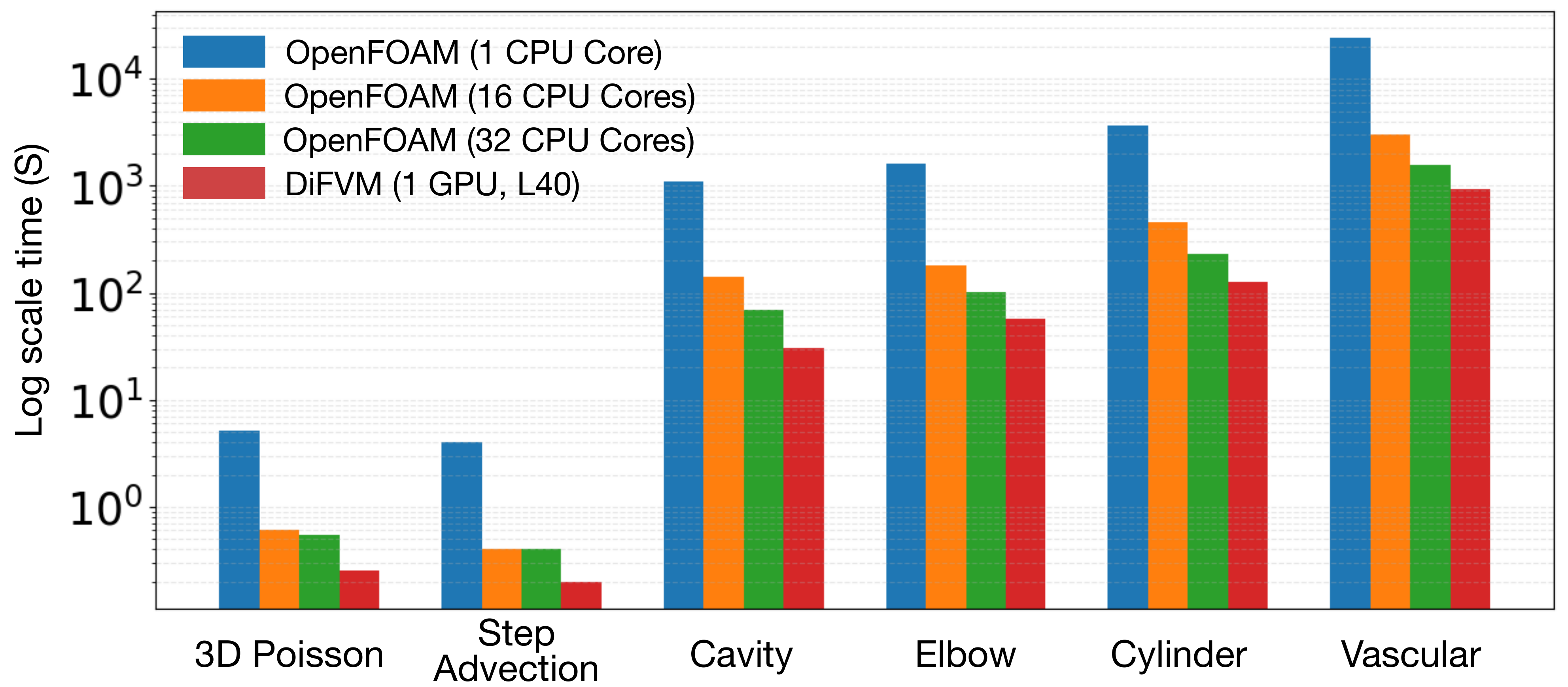}
\caption{Wall-clock time comparison between DiFVM (1 Nvidia L40 GPU) and OpenFOAM (1, 16, and 32 CPU cores) across all forward cases on a logarithmic scale. DiFVM consistently outperforms all CPU configurations, delivering $16\text{--}33\times$ speedup over single-core and $1.6\text{--}2.3\times$ over 32-core parallel OpenFOAM across all cases.}
\label{fig:speed}
\end{figure}

Figure~\ref{fig:speed} reports the wall-clock time of DiFVM across all forward benchmark cases, compared against OpenFOAM executed on a single CPU core, 16 CPU cores, and 32 CPU cores, with DiFVM running on an NVIDIA L40 GPU. All comparisons use identical meshes, boundary conditions, and numerical settings.
It is important to note that all DiFVM timings are obtained on a \emph{single} NVIDIA L40 GPU, without any multi-GPU or multi-node parallelism. The speedup profile exhibits a characteristic trend with problem size that reflects the interplay between GPU vectorization efficiency, memory bandwidth, and arithmetic intensity. For the smallest problems (3D Poisson and step advection), a single GPU already delivers ${\sim}16\times$ speedup over single-core CPU due to XLA kernel fusion and elimination of Python-level loop overhead, though the workload is insufficient to fully saturate the GPU's compute units. The largest relative speedups are observed for medium-to-large transient problems, which achieve ${\sim}25\text{--}33\times$ over single-core CPU and ${\sim}2\times$ over 32-core parallel OpenFOAM, where the computational workload is large enough to amortize GPU kernel launch overhead while remaining within the bandwidth-efficient operating regime. Critically, a single GPU remains ${\sim}1.4\times$ faster than 32-core parallel OpenFOAM for all cases, underscoring the efficiency of DiFVM's graph-based vectorization strategy on a single accelerator.

These gains reflect DiFVM's two-level vectorization strategy. Spatially, all finite-volume flux computations are reformulated as batched gather--compute--scatter operations over the static mesh graph and compiled into fused XLA GPU kernels, eliminating sequential cell-by-cell traversal. Temporally, the PISO time-marching loop is unrolled via \texttt{jax.lax.scan} into a single static computation graph, which XLA compiles to a small number of GPU executables with no host--device synchronization overhead between time steps. Together, these transformations ensure that even the largest transient benchmarks, involving thousands of time steps and iterative pressure solves, run almost entirely on-device without Python overhead.

The current benchmarks represent only a fraction of the hardware capabilities available on modern accelerators. The 3D vascular benchmark (${\sim} 10^6$ cells) occupies approximately 2.3~GB of GPU memory in forward-only mode, leaving substantial headroom on the NVIDIA L40 (48~GB). This indicates that meshes one to two orders of magnitude larger could be handled without architectural changes, suggesting that DiFVM's performance advantages will become even more pronounced as problem size scales toward the full capacity of the hardware. For reverse-mode AD through long transient simulations, memory is managed via gradient checkpointing, which reduces the memory footprint at the cost of one additional forward pass, enabling differentiable simulation at scales that would otherwise exceed GPU memory capacity.

\subsection{Limitations and Future Work}
\label{sec:limitations}

The current framework solves the incompressible Navier--Stokes equations for rigid-wall geometries. High-fidelity extensions including FSI, turbulence modeling via large-eddy simulation (LES) or Reynolds-averaged Navier--Stokes (RANS) closures, and multiscale coupling with reduced-order 1D/0D cardiovascular network models are not yet incorporated. These capabilities are essential for patient-specific hemodynamic simulation, particularly for applications involving compliant vessel walls, turbulent aortic flow, or systemic circulation modeling. Extending DiFVM to FSI and multi-physics coupling, building on the differentiable FSI framework of Diff-FlowFSI~\cite{fan2026diff}, is a primary direction for future work.

The current implementation executes on a single GPU. While the largest benchmark considered here leaves substantial headroom on a single NVIDIA L40 (48~GB), production-scale simulations, e.g., patient-specific vascular networks, high-resolution LES, or full-body hemodynamic models, will require meshes of $10^8$--$10^9$ cells that exceed single-GPU capacity. DiFVM's graph-based formulation is architecturally compatible with multi-GPU execution: the static mesh graph $\mathcal{G}$ can be partitioned across devices using standard graph partitioning algorithms (e.g., METIS~\cite{karypis1998fast}), with inter-partition halo exchange handled via MPI or JAX's native \texttt{jax.distributed} interface. Under this multi-GPU paradigm, each device owns a subgraph of $\mathcal{G}$ and executes the same scatter/gather message-passing kernels locally, with boundary face fluxes communicated across partitions. Because the graph structure is static, the communication pattern is fixed at initialization and can be optimized once at compile time. Extending DiFVM to multi-GPU and multi-node execution via MPI--GPU parallelism is a primary scalability direction, and would unlock the full potential of differentiable unstructured-mesh CFD at production engineering scales, at which the performance gap over CPU-based solvers is expected to widen further.

\section{Conclusion}
\label{sec:conclusion}

We presented DiFVM, the first GPU-accelerated, end-to-end differentiable finite-volume CFD solver operating natively on unstructured polyhedral meshes. The central contribution is not an incremental improvement to existing solvers, but an architectural reformulation: by establishing a structural isomorphism between FVM discretization and graph neural network message-passing, DiFVM transforms the irregular connectivity of unstructured meshes, historically the primary obstacle to GPU vectorization, into a static compile-time data structure amenable to fully fused XLA GPU kernels. This reformulation is exact and lossless; every flux computation, PISO pressure--velocity coupling iteration, and Rhie--Chow correction is expressed as a scatter/gather operation on the mesh graph without any approximation to the underlying numerical method. End-to-end automatic differentiation (AD) through the complete simulation pipeline is provided by JAX's reverse-mode AD, with gradient checkpointing and implicit differentiation through the Poisson solve ensuring memory efficiency for long transient simulations.

Forward validation across six benchmarks of increasing geometric and physical complexity, from canonical flows to patient-specific cardiovascular hemodynamics, establishes quantitative agreement with OpenFOAM, with relative errors consistently below $\mathcal{O}(10^{-5})$ across all field quantities and geometries. On a single NVIDIA L40 GPU, DiFVM consistently outperforms 32-core parallel OpenFOAM across all cases, with the performance advantage growing with problem complexity. These results are achieved without any multi-GPU parallelism; the static graph structure of DiFVM is architecturally compatible with domain decomposition across multiple devices, suggesting that the performance advantage over CPU-based solvers will widen further at production engineering scales.

End-to-end differentiability is demonstrated on two inverse problems of increasing complexity: scalar boundary parameter recovery in cavity flow, and simultaneous inference of six Windkessel parameters in pulsatile bifurcation flow, with all parameters converging to within $1\%$ of ground truth from sparse pressure and flow-rate observations. These results show that the gradient propagates correctly through thousands of time steps, iterative solves, and nonlinear flux computations, confirming the correctness of the AD pipeline across the full complexity of a transient CFD simulation.

DiFVM addresses a structural gap in the differentiable CFD ecosystem: the absence of a GPU-native, gradient-capable solver on the unstructured meshes that define the most consequential engineering and biomedical applications. By unifying FVM discretization, GPU-native execution, and end-to-end differentiability within a single JAX-based framework, DiFVM provides the missing computational substrate for gradient-driven workflows including PDE-constrained optimization, Bayesian parameter inference, data assimilation, and in-the-loop training of hybrid ML-CFD modeling on geometrically complex domains that were previously inaccessible to differentiable simulation.

\bibliographystyle{elsarticle-num}
\bibliography{ref,myref}

@article{akhare2025implicit,
title = {Implicit neural differentiable model for spatiotemporal dynamics},
journal = {Computer Methods in Applied Mechanics and Engineering},
volume = {446},
pages = {118280},
year = {2025},
issn = {0045-7825},
doi = {https://doi.org/10.1016/j.cma.2025.118280},
url = {https://www.sciencedirect.com/science/article/pii/S0045782525005523},
author = {Deepak Akhare and Pan Du and Tengfei Luo and Jian-Xun Wang},
}

@article{akhare2025hybridndiff,
  title={HybridNDiff-UQ: Uncertainty Quantification for Hybrid Neural Differentiable Modeling},
  author={Akhare, Deepak and Luo, Tengfei and Wang, Jian-Xun},
  journal={Theoretical and Applied Mechanics Letters},
  pages={100609},
  year={2025},
  publisher={Elsevier}
}

@article{shang2025jax,
  title={JAX-BTE: a GPU-accelerated differentiable solver for phonon Boltzmann transport equations},
  author={Shang, Wenjie and Zhou, Jiahang and Panda, JP and Xu, Zhihao and Liu, Yi and Du, Pan and Wang, Jian-Xun and Luo, Tengfei},
  journal={npj Computational Materials},
  volume={11},
  number={1},
  pages={1--12},
  year={2025},
  publisher={Nature Publishing Group}
}

@article{fan2025neural,
  title={Neural differentiable modeling with diffusion-based super-resolution for two-dimensional spatiotemporal turbulence},
  author={Fan, Xiantao and Akhare, Deepak and Wang, Jian-Xun},
  journal={Computer Methods in Applied Mechanics and Engineering},
  volume={433},
  pages={117478},
  year={2025},
  publisher={Elsevier}
}

@article{akhare2024probabilistic,
  title={Probabilistic physics-integrated neural differentiable modeling for isothermal chemical vapor infiltration process},
  author={Akhare, Deepak and Chen, Zeping and Gulotty, Richard and Luo, Tengfei and Wang, Jian-Xun},
  journal={npj Computational Materials},
  volume={10},
  number={1},
  pages={120},
  year={2024},
  publisher={Nature Publishing Group UK London}
}

@article{liu2024multi,
	title={Multi-resolution partial differential equations preserved learning framework for spatiotemporal dynamics},
	author={Liu, Xin-Yang and Zhu, Min and Lu, Lu and Sun, Hao and Wang, Jian-Xun},
	journal={Communications Physics},
	volume={7},
	number={1},
	pages={31},
	year={2024},
	publisher={Nature Publishing Group UK London}
}

@article{fan2024differentiable,
	title={Differentiable hybrid neural modeling for fluid-structure interaction},
	author={Fan, Xiantao and Wang, Jian-Xun},
	journal={Journal of Computational Physics},
	volume={496},
	pages={112584},
	year={2024},
	publisher={Elsevier}
}

@article{wang2019data,
	title={Data-augmented modeling of intracranial pressure},
	author={Wang, Jian-Xun and Hu, Xiao and Shadden, Shawn C},
	journal={Annals of biomedical engineering},
	volume={47},
	number={3},
	pages={714--730},
	year={2019},
	publisher={Springer}
}

@article{wang2019prediction,
	title={Prediction of Reynolds stresses in high-Mach-number turbulent boundary layers using physics-informed machine learning},
	author={Wang, Jian-Xun and Huang, Junji and Duan, Lian and Xiao, Heng},
	journal={Theoretical and Computational Fluid Dynamics},
	volume={33},
	number={1},
	pages={1--19},
	year={2019},
	publisher={Springer}
}

@article{yang2019predictive,
	title={Predictive large-eddy-simulation wall modeling via physics-informed neural networks},
	author={Yang, XIA and Zafar, S and Wang, J-X and Xiao, H},
	journal={Physical Review Fluids},
	volume={4},
	number={3},
	pages={034602},
	year={2019},
	publisher={APS}
}

@article{wang2017physics,
	title={Physics-informed machine learning approach for reconstructing Reynolds stress modeling discrepancies based on {DNS} data},
	author={Wang, Jian-Xun and Wu, Jin-Long and Xiao, Heng},
	journal={Physical Review Fluids},
	volume={2},
	number={3},
	pages={034603},
	year={2017},
	publisher={APS}
}

@article{wang2016data,
	title={Data-driven CFD modeling of turbulent flows through complex structures},
	author={Wang, Jian-Xun and Xiao, Heng},
	journal={International Journal of Heat and Fluid Flow},
	volume={62},
	pages={138--149},
	year={2016},
	publisher={Elsevier}
}

@article{akhare2023diffhybrid,
  title={Diffhybrid-uq: uncertainty quantification for differentiable hybrid neural modeling},
  author={Akhare, Deepak and Luo, Tengfei and Wang, Jian-Xun},
  journal={arXiv preprint arXiv:2401.00161},
  year={2023}
}

@article{mader2008adjoint,
  title={ADjoint: An approach for the rapid development of discrete adjoint solvers},
  author={Mader, Charles A and Martins, Joaquim RRA and Alonso, Juan J and Van Der Weide, Edwin},
  journal={AIAA journal},
  volume={46},
  number={4},
  pages={863--873},
  year={2008}
}

@article{karypis1998fast,
  title   = {A fast and high quality multilevel scheme for partitioning
             irregular graphs},
  author  = {Karypis, George and Kumar, Vipin},
  journal = {SIAM Journal on Scientific Computing},
  volume  = {20},
  number  = {1},
  pages   = {359--392},
  year    = {1998}
}

@article{weller1998tensorial,
  title={A tensorial approach to computational continuum mechanics using object-oriented techniques},
  author={Weller, Henry G and Tabor, Gavin and Jasak, Hrvoje and Fureby, Christer},
  journal={Computers in physics},
  volume={12},
  number={6},
  pages={620--631},
  year={1998},
  publisher={American Institute of Physics}
}

@article{updegrove2017simvascular,
  title={SimVascular: an open source pipeline for cardiovascular simulation},
  author={Updegrove, Adam and Wilson, Nathan M and Merkow, Jameson and Lan, Hongzhi and Marsden, Alison L and Shadden, Shawn C},
  journal={Annals of biomedical engineering},
  volume={45},
  number={3},
  pages={525--541},
  year={2017},
  publisher={Springer}
}

@article{blondel2024elements,
  title={The elements of differentiable programming},
  author={Blondel, Mathieu and Roulet, Vincent},
  journal={arXiv preprint arXiv:2403.14606},
  year={2024}
}

@article{issa1986solution,
  title={Solution of the implicitly discretised fluid flow equations by operator-splitting},
  author={Issa, Raad I},
  journal={Journal of computational physics},
  volume={62},
  number={1},
  pages={40--65},
  year={1986},
  publisher={Elsevier}
}

@phdthesis{jasak1996error,
  author  = {Jasak, Hrvoje},
  title   = {Error Analysis and Estimation for the Finite Volume Method with Applications to Fluid Flows},
  school  = {Imperial College London},
  year    = {1996},
}

@article{rhie1983numerical,
  title={Numerical study of the turbulent flow past an airfoil with trailing edge separation},
  author={Rhie, Chae M and Chow, Wei-Liang},
  journal={AIAA journal},
  volume={21},
  number={11},
  pages={1525--1532},
  year={1983}
}

@article{paszke2019pytorch,
  title={Pytorch: An imperative style, high-performance deep learning library},
  author={Paszke, Adam and Gross, Sam and Massa, Francisco and Lerer, Adam and Bradbury, James and Chanan, Gregory and Killeen, Trevor and Lin, Zeming and Gimelshein, Natalia and Antiga, Luca and others},
  journal={Advances in neural information processing systems},
  volume={32},
  year={2019}
}

@inproceedings{abadi2016tensorflow,
  title={$\{$TensorFlow$\}$: a system for $\{$Large-Scale$\}$ machine learning},
  author={Abadi, Mart{\'\i}n and Barham, Paul and Chen, Jianmin and Chen, Zhifeng and Davis, Andy and Dean, Jeffrey and Devin, Matthieu and Ghemawat, Sanjay and Irving, Geoffrey and Isard, Michael and others},
  booktitle={12th USENIX symposium on operating systems design and implementation (OSDI 16)},
  pages={265--283},
  year={2016}
}

@article{shankar2025differentiable,
  title={Differentiable turbulence: Closure as a partial differential equation constrained optimization},
  author={Shankar, Varun and Chakraborty, Dibyajyoti and Viswanathan, Venkatasubramanian and Maulik, Romit},
  journal={Physical Review Fluids},
  volume={10},
  number={2},
  pages={024605},
  year={2025},
  publisher={APS}
}

@article{beck2019deep,
  title={Deep neural networks for data-driven LES closure models},
  author={Beck, Andrea and Flad, David and Munz, Claus-Dieter},
  journal={Journal of Computational Physics},
  volume={398},
  pages={108910},
  year={2019},
  publisher={Elsevier}
}

@article{sirignano2020dpm,
  title={DPM: A deep learning PDE augmentation method with application to large-eddy simulation},
  author={Sirignano, Justin and MacArt, Jonathan F and Freund, Jonathan B},
  journal={Journal of Computational Physics},
  volume={423},
  pages={109811},
  year={2020},
  publisher={Elsevier}
}

@article{fang2023toward,
  title={Toward more general turbulence models via multicase computational-fluid-dynamics-driven training},
  author={Fang, Yuan and Zhao, Yaomin and Waschkowski, Fabian and Ooi, Andrew SH and Sandberg, Richard D},
  journal={AIAA Journal},
  volume={61},
  number={5},
  pages={2100--2115},
  year={2023},
  publisher={American Institute of Aeronautics and Astronautics}
}

@article{macart2021embedded,
  title={Embedded training of neural-network subgrid-scale turbulence models},
  author={MacArt, Jonathan F and Sirignano, Justin and Freund, Jonathan B},
  journal={Physical Review Fluids},
  volume={6},
  number={5},
  pages={050502},
  year={2021},
  publisher={APS}
}

@article{wu2019reynolds,
  title={Reynolds-averaged Navier--Stokes equations with explicit data-driven Reynolds stress closure can be ill-conditioned},
  author={Wu, Jinlong and Xiao, Heng and Sun, Rui and Wang, Qiqi},
  journal={Journal of Fluid Mechanics},
  volume={869},
  pages={553--586},
  year={2019},
  publisher={Cambridge University Press}
}

@article{prakash2024invariant,
  title={Invariant data-driven subgrid stress modeling on anisotropic grids for large eddy simulation},
  author={Prakash, Aviral and Jansen, Kenneth E and Evans, John A},
  journal={Computer Methods in Applied Mechanics and Engineering},
  volume={422},
  pages={116807},
  year={2024},
  publisher={Elsevier}
}

@article{akolekar2019development,
  title={Development and use of machine-learnt algebraic Reynolds stress models for enhanced prediction of wake mixing in low-pressure turbines},
  author={Akolekar, Harshal D and Weatheritt, Jack and Hutchins, Nicholas and Sandberg, Richard D and Laskowski, Gregory and Michelassi, Vittorio},
  journal={Journal of Turbomachinery},
  volume={141},
  number={4},
  pages={041010},
  year={2019},
  publisher={American Society of Mechanical Engineers}
}

@article{pan2018data,
  title={Data-driven discovery of closure models},
  author={Pan, Shaowu and Duraisamy, Karthik},
  journal={SIAM Journal on Applied Dynamical Systems},
  volume={17},
  number={4},
  pages={2381--2413},
  year={2018},
  publisher={SIAM}
}

@article{duraisamy2019turbulence,
  title={Turbulence modeling in the age of data},
  author={Duraisamy, Karthik and Iaccarino, Gianluca and Xiao, Heng},
  journal={Annual review of fluid mechanics},
  volume={51},
  number={1},
  pages={357--377},
  year={2019},
  publisher={Annual Reviews}
}

@article{maulik2022efficient,
  title={Efficient high-dimensional variational data assimilation with machine-learned reduced-order models},
  author={Maulik, Romit and Rao, Vishwas and Wang, Jiali and Mengaldo, Gianmarco and Constantinescu, Emil and Lusch, Bethany and Balaprakash, Prasanna and Foster, Ian and Kotamarthi, Rao},
  journal={Geoscientific Model Development},
  volume={15},
  number={8},
  pages={3433--3445},
  year={2022},
  publisher={Copernicus GmbH}
}

@article{da2018ensemble,
  title={Ensemble-based state estimator for aerodynamic flows},
  author={da Silva, Andre FC and Colonius, Tim},
  journal={AIAA Journal},
  volume={56},
  number={7},
  pages={2568--2578},
  year={2018},
  publisher={American Institute of Aeronautics and Astronautics}
}

@article{wang2025variational,
  title={Variational data assimilation in wall turbulence: from outer observations to wall stress and pressure},
  author={Wang, Mengze and Zaki, Tamer A},
  journal={Journal of Fluid Mechanics},
  volume={1008},
  pages={A26},
  year={2025},
  publisher={Cambridge University Press}
}

@article{brunton2020machine,
  title={Machine learning for fluid mechanics},
  author={Brunton, Steven L and Noack, Bernd R and Koumoutsakos, Petros},
  journal={Annual review of fluid mechanics},
  volume={52},
  number={1},
  pages={477--508},
  year={2020},
  publisher={Annual Reviews}
}

@article{li2019data,
  title={Data-driven constraint approach to ensure low-speed performance in transonic aerodynamic shape optimization},
  author={Li, Jichao and He, Sicheng and Martins, Joaquim RRA},
  journal={Aerospace Science and Technology},
  volume={92},
  pages={536--550},
  year={2019},
  publisher={Elsevier}
}

@article{bui2004aerodynamic,
  title={Aerodynamic data reconstruction and inverse design using proper orthogonal decomposition},
  author={Bui-Thanh, Tan and Damodaran, Murali and Willcox, Karen},
  journal={AIAA journal},
  volume={42},
  number={8},
  pages={1505--1516},
  year={2004}
}

@software{jax2018github,
  author = {James Bradbury and Roy Frostig and Peter Hawkins and Matthew James Johnson and Chris Leary and Dougal Maclaurin and George Necula and Adam Paszke and Jake Vander{P}las and Skye Wanderman-{M}ilne and Qiao Zhang},
  title = {{JAX}: composable transformations of {P}ython+{N}um{P}y programs},
  url = {http://github.com/jax-ml/jax},
  version = {0.3.13},
  year = {2018},
}

@inproceedings{pfaff2020learning,
  title={Learning mesh-based simulation with graph networks},
  author={Pfaff, Tobias and Fortunato, Meire and Sanchez-Gonzalez, Alvaro and Battaglia, Peter},
  booktitle={International conference on learning representations},
  year={2020}
}

@article{bar2019learning,
  title={Learning data-driven discretizations for partial differential equations},
  author={Bar-Sinai, Yohai and Hoyer, Stephan and Hickey, Jason and Brenner, Michael P},
  journal={Proceedings of the National Academy of Sciences},
  volume={116},
  number={31},
  pages={15344--15349},
  year={2019},
  publisher={National Academy of Sciences}
}

@article{kochkov2021machine,
  title={Machine learning--accelerated computational fluid dynamics},
  author={Kochkov, Dmitrii and Smith, Jamie A and Alieva, Ayya and Wang, Qing and Brenner, Michael P and Hoyer, Stephan},
  journal={Proceedings of the National Academy of Sciences},
  volume={118},
  number={21},
  pages={e2101784118},
  year={2021},
  publisher={National Academy of Sciences}
}

@inproceedings{holl2024bf,
  title={$\Phi$-Flow: Differentiable Simulations for PyTorch, TensorFlow and Jax},
  author={Holl, Philipp and Thuerey, Nils},
  booktitle={Forty-first international conference on machine learning},
  year={2024}
}

@article{bezgin2025jax,
  title={JAX-Fluids 2.0: towards HPC for differentiable CFD of compressible two-phase flows},
  author={Bezgin, Deniz A and Buhendwa, Aaron B and Adams, Nikolaus A},
  journal={Computer Physics Communications},
  volume={308},
  pages={109433},
  year={2025},
  publisher={Elsevier}
}

@article{weymouth2024waterlily,
  title={WaterLily. jl: A differentiable and backend-agnostic Julia solver to simulate incompressible viscous flow and dynamic bodies},
  author={Weymouth, Gabriel D and Font, Bernat},
  journal={arXiv preprint arXiv:2407.16032},
  year={2024}
}

@article{franz2025pict,
  title={PICT--A Differentiable, GPU-Accelerated Multi-Block PISO Solver for Simulation-Coupled Learning Tasks in Fluid Dynamics},
  author={Franz, Aleksandra and Wei, Hao and Guastoni, Luca and Thuerey, Nils},
  journal={arXiv preprint arXiv:2505.16992},
  year={2025}
}

@article{economon2016su2,
  title={SU2: An open-source suite for multiphysics simulation and design},
  author={Economon, Thomas D and Palacios, Francisco and Copeland, Sean R and Lukaczyk, Trent W and Alonso, Juan J},
  journal={Aiaa Journal},
  volume={54},
  number={3},
  pages={828--846},
  year={2016},
  publisher={American Institute of Aeronautics and Astronautics}
}

@article{mitusch2019dolfin,
  title={dolfin-adjoint 2018.1: automated adjoints for FEniCS and Firedrake},
  author={Mitusch, Sebastian and Funke, Simon and Dokken, J{\o}rgen},
  journal={Journal of Open Source Software},
  volume={4},
  number={38},
  pages={1292},
  year={2019},
  publisher={The Open Journal}
}

@incollection{ccmSu,
author = "Siemens Digital Industries Software",
title = "Simcenter \uppercase{STAR-CCM+} {U}ser {G}uide, version 2021.1",
editor = "",
booktitle = "Adaptive Mesh Refinement for Overset Meshes",
publisher = "Siemens",
address = "",
year = "2021",
pages = "3067-3070"
}

@article{zhuang2021learned,
  title={Learned discretizations for passive scalar advection in a two-dimensional turbulent flow},
  author={Zhuang, Jiawei and Kochkov, Dmitrii and Bar-Sinai, Yohai and Brenner, Michael P and Hoyer, Stephan},
  journal={Physical Review Fluids},
  volume={6},
  number={6},
  pages={064605},
  year={2021},
  publisher={APS}
}

@article{bezgin2021data,
  title={A data-driven physics-informed finite-volume scheme for nonclassical undercompressive shocks},
  author={Bezgin, Deniz A and Schmidt, Steffen J and Adams, Nikolaus A},
  journal={Journal of Computational Physics},
  volume={437},
  pages={110324},
  year={2021},
  publisher={Elsevier}
}

@article{list2022learned,
  title={Learned turbulence modelling with differentiable fluid solvers: physics-based loss functions and optimisation horizons},
  author={List, Bj{\"o}rn and Chen, Li-Wei and Thuerey, Nils},
  journal={Journal of Fluid Mechanics},
  volume={949},
  pages={A25},
  year={2022},
  publisher={Cambridge University Press}
}

@article{um2020solver,
  title={Solver-in-the-loop: Learning from differentiable physics to interact with iterative pde-solvers},
  author={Um, Kiwon and Brand, Robert and Fei, Yun Raymond and Holl, Philipp and Thuerey, Nils},
  journal={Advances in neural information processing systems},
  volume={33},
  pages={6111--6122},
  year={2020}
}

@article{lino2023current,
  title={Current and emerging deep-learning methods for the simulation of fluid dynamics},
  author={Lino, Mario and Fotiadis, Stathi and Bharath, Anil A and Cantwell, Chris D},
  journal={Proceedings of the Royal Society A: Mathematical, Physical and Engineering Sciences},
  volume={479},
  number={2275},
  year={2023},
  publisher={The Royal Society}
}

@inproceedings{belbute2020combining,
  title={Combining differentiable PDE solvers and graph neural networks for fluid flow prediction},
  author={Belbute-Peres, Filipe De Avila and Economon, Thomas and Kolter, Zico},
  booktitle={international conference on machine learning},
  pages={2402--2411},
  year={2020},
  organization={PMLR}
}

@article{fan2026diff,
  title={Diff-FlowFSI: A GPU-optimized differentiable CFD platform for high-fidelity turbulence and FSI simulations},
  author={Fan, Xiantao and Liu, Xin-Yang and Wang, Meng and Wang, Jian-Xun},
  journal={Computer Methods in Applied Mechanics and Engineering},
  volume={448},
  pages={118455},
  year={2026},
  publisher={Elsevier}
}

@inproceedings{zawawi2018review,
  title={A review: Fundamentals of computational fluid dynamics (CFD)},
  author={Zawawi, Mohd Hafiz and Saleha, A and Salwa, A and Hassan, NH and Zahari, Nazirul Mubin and Ramli, Mohd Zakwan and Muda, Zakaria Che},
  booktitle={AIP conference proceedings},
  volume={2030},
  number={1},
  pages={020252},
  year={2018},
  organization={AIP Publishing LLC}
}

@incollection{jujjavarapu2024computational,
  title={Computational Fluid Dynamics in Biomedical Engineering},
  author={Jujjavarapu, Satya Eswari and Kumar, Tukendra and Gupta, Sharda},
  booktitle={Computational Fluid Dynamics Applications in Bio and Biomedical Processes: Biotechnology Applications},
  pages={101--125},
  year={2024},
  publisher={Springer}
}

@article{fujii2005progress,
  title={Progress and future prospects of CFD in aerospace—Wind tunnel and beyond},
  author={Fujii, Kozo},
  journal={Progress in Aerospace Sciences},
  volume={41},
  number={6},
  pages={455--470},
  year={2005},
  publisher={Elsevier}
}

@article{rojas2016review,
  title={A review of the computational fluid dynamics simulation software: Advantages, disadvantages and main applications},
  author={Rojas-Sola, Jos{\'e} Ignacio and Garc{\'\i}a-Baena, Carlos and Hermoso-Orz{\'a}ez, Manuel Jes{\'u}s},
  journal={Journal of Magnetohydrodynamics and Plasma Research},
  volume={21},
  number={4},
  pages={417--424},
  year={2016},
  publisher={Nova Science Publishers, Inc.}
}

@article{pant2017inverse,
  title={Inverse problems in reduced order models of cardiovascular haemodynamics: aspects of data assimilation and heart rate variability},
  author={Pant, Sanjay and Corsini, Chiara and Baker, Catriona and Hsia, Tain-Yen and Pennati, Giancarlo and Vignon-Clementel, Irene E},
  journal={Journal of The Royal Society Interface},
  volume={14},
  number={126},
  pages={20160513},
  year={2017},
  publisher={The Royal Society}
}

@article{sapienza2024differentiable,
  title={Differentiable programming for differential equations: A review},
  author={Sapienza, Facundo and Bolibar, Jordi and Sch{\"a}fer, Frank and Groenke, Brian and Pal, Avik and Boussange, Victor and Heimbach, Patrick and Hooker, Giles and P{\'e}rez, Fernando and Persson, Per-Olof and others},
  journal={arXiv preprint arXiv:2406.09699},
  year={2024}
}

@article{Julia-2017,
    title={Julia: A fresh approach to numerical computing},
    author={Bezanson, Jeff and Edelman, Alan and Karpinski, Stefan and Shah, Viral B},
    journal={SIAM {R}eview},
    volume={59},
    number={1},
    pages={65--98},
    year={2017},
    publisher={SIAM},
    doi={10.1137/141000671},
    url={https://epubs.siam.org/doi/10.1137/141000671}
}

@article{posey2022gpu,
  title={GPU-based HPC and AI developments for CFD},
  author={Posey, S and Luitjens, J and Hennigh, O and Oberlin, S},
  journal={Maui, Hawaii, USA, July},
  pages={11--15},
  year={2022}
}

@article{Wilson2013,
  title = {The Vascular Model Repository: A Public Resource of Medical Imaging Data and Blood Flow Simulation Results},
  author = {Wilson, Nathan M. and Ortiz, Ana K. and Johnson, Allison B.},
  year = {2013},
  journal = {Journal of Medical Devices},
  volume = {7},
  number = {4},
  pages = {0409231--409231}
}

@article{westerhof2009arterial,
  title={The arterial windkessel},
  author={Westerhof, Nico and Lankhaar, Jan-Willem and Westerhof, Berend E},
  journal={Medical \& biological engineering \& computing},
  volume={47},
  number={2},
  pages={131--141},
  year={2009},
  publisher={Springer}
}
\clearpage


\end{document}